\documentclass[12pt,preprint]{aastex}

\usepackage{epsfig}
\begin{document}

\title{Comprehensive study of the X-ray flares from gamma-ray bursts observed by {\em Swift}}
\author{Shuang-Xi Yi$^{1,2}$, Shao-Qiang Xi$^{4}$, Hai Yu$^{2,3}$, F. Y. Wang$^{2,3}$, Hui-Jun Mu$^{5}$, Lian-Zhong L\"u$^{6, 7}$ \& En-Wei Liang$^{6,7}$}
\affil{$^{1}$College of Physics and Engineering, Qufu Normal University, Qufu 273165, China; \\
       $^{2}$School of Astronomy and Space Science, Nanjing University, Nanjing 210093, China; fayinwang@nju.edu.cn\\
       $^{3}$Key laboratory of Modern Astronomy and Astrophysics (Nanjing University), Nanjing 210093, China\\
       $^{4}$Department of Mathematics and Physics, Officers College of CAPF, Chengdu, 610213, China; \\
       $^{5}$Department of Astronomy and Institute of Theoretical Physics and Astrophysics, Xiamen University, Xiamen, Fujian 361005, China; \\
       $^{6}$GXU-NAOC Center for Astrophysics and Space Sciences, Department of Physics, Guangxi University, Nanning 530004; lew@gxu.edu.cn\\
       $^{7}$Guangxi Key Laboratory for the Relativistic Astrophysics, Nanning 530004, China. \\
       }

\begin{abstract}
X-ray flares are generally supposed to be produced by the later
central engine activities, and may share the similar physical origin
with prompt emission of gamma-ray bursts (GRBs). In this paper, we
have analyzed all significant X-ray flares from the GRBs observed by
{\em Swift} from April 2005 to March 2015. The catalog contains 468
bright X-ray flares, including 200 flares with redshifts. We obtain
the fitting results of X-ray flares, such as start time, peak time,
duration, peak flux, fluence, peak luminosity, and mean luminosity.
The peak luminosity decreases with peak time, following a power-law
behavior $L_p \propto T_{peak,z}^{-1.27}$. The flare duration
increases with peak time. The 0.3-10 keV isotropic energy of X-ray
flares distribution is a lognormal peaked at $10^{51.2}$ erg. We
also study the frequency distributions of flare parameters,
including energies, durations, peak fluxes, rise times, decay times
and waiting times. Power-law distributions of energies, durations,
peak fluxes, and waiting times are found in GRB X-ray flares and
solar flares. These distributions could be well explained by a
fractal-diffusive, self-organized criticality model. Some
theoretical models basing on magnetic reconnection have been
proposed to explain X-ray flares. Our result shows that the
relativistic jets of GRBs may be Poynting-flux dominated.
\end{abstract}

\section{Introduction}
X-ray flare is one of the most common phenomena in the afterglow
phase of gamma-ray burst (GRBs) in the {\em Swift} satellite era
(Burrows et al. 2005; Falcone et al. 2006; Zhang et al. 2006; Nousek
et al. 2006). {There are about one-third of {\em Swift} GRBs
with remarkable X-ray flares. X-ray flares have been observed in
both long and short GRBs (Romano et al. 2006; Falcone et al. 2006;
Campana et al. 2006; Margutti et al. 2011). Generally, the X-ray
flare shows sharp rise and sharp decay. They usually happens at
$10^2-10^5$ s after the prompt emission (Falcone et al. 2007;
Chincarini et al. 2007, 2010). The fluences of most X-ray flares are
smaller than the prompt emission observed by {\em Swift/BAT}. Their
average fluence is about 10 percent of the prompt emission
statistically (Falcone et al. 2007; Chincarini et al. 2010). From
the temporal behavior and spectral property, it is believed that the
X-ray flare is from a distinct emission mechanism, which is
different from the underlying afterglow emission. While the temporal
behavior of flares is very similar to that of prompt emission
pulses. Therefore, X-ray flares may have the same physical origin as
the prompt pulses of GRBs (Burrows et al. 2005; Falcone et al. 2006,
2007; Zhang et al. 2006; Nousek et al. 2006; Liang et al. 2006;
Chincarini et al. 2007, 2010; Hou et al. 2013; Wu et al. 2013; Yi et
al. 2015a). Both X-ray flares and GRBs are powered by the central
engine activities, therefore the properties of X-ray flares can
provide an important clue to understand the mechanism of GRB
phenomenon. Some theoretical models have been proposed, such as
fragmentation of the collapsing star (King et al. 2005) ,
fragmentation of the accretion disk (Perna et al. 2006),
intermittent accretion behavior caused by a time variable magnetic
barrier (Proga \& Zhang 2006), magnetic reconnection from a
post-merger millisecond pulsar (Dai et al. 2006), and magnetic
dissipation in a decelerating shell (Giannios 2006).

Some surveys on the X-ray flares of GRBs observed by {\em Swift}
have been carried out. The studies by Falcone et al. (2007) and
Chincarini et al. (2007) selected dozens of flares in the early
period of {\em Swift}. They fitted the X-ray flare with a broken
power-law or multiple broken power-law functions, and obtained the
fitting parameters of X-ray flares. These studies suggested that
flares are produced by late activities of central engine. Follow-up
studies with new sample of flares indicated that X-ray flares have
some correlations among the flare's parameters (Chincarini et al.
2010; Bernardini et al. 2011; Swenson \& Roming 2014). They
confirmed that the late-time internal dissipation origin seems the
most promising explanation for flares. Margutti et al. (2011)
studied X-ray flare candidates in short GRBs, and found that short
GRB flares show similar observational properties of long ones after
accounting for the central engine time-scales and energy budget.
Besides that, Li et al. (2012) also investigated 24 optical flares
from 19 GRBs. They suggested that, similar to the X-ray flares, the
optical flares are related to the erratic behavior of the central
engine. Guidorzi et al. (2015) found that the waiting time
distributions of prompt pulses and X-ray flares show a similar
power-law behavior.

X-ray flares are the common astrophysical phenomena throughout the
universe. Interestingly, Wang \& Dai (2013) compiled 83 GRB flares
and 11595 solar hard X-ray flares from {\em RHESSI} during 2002-2007
and performed a statistical comparison between them. They found the
energy, duration, and waiting-time distributions of X-ray flares are
similar to those of solar flares, which suggest a similar physical
origin of the both events. This result is supported by later
numerical simulations (Harko et al. 2015). Harko et al. (2015)
numerically investigated the possibility that self-organized
criticality (SOC) appears in a one-dimensional magnetized flow,
which can be applied to GRB X-ray flares. Wang et al. (2015) studied
the energy, duration and waiting time distributions of X-ray flares
from Swift J1644+57 (Burrows et al. 2011), Galactic center black
hole Sgr A$^*$ (Neilsen et al. 2013), and M87 (Harris et al. 2009;
Abramowski et al. 2012). These distributions of X-ray flares in
different systems show similar power-law distributions. So X-ray
flares from astrophysical systems with spatial and mass scales
different by many orders of magnitudes show similar behavior, which
may indicate that they have similar physical origin.

In this paper, we analyse the ten-year X-ray flare data of {\em
Swift/XRT} until the end of March 2015, and study the distributions
of energy, duration, waiting time, rise time, decay time, peak time
and peak flux. This paper is organized as follows. In Section 2, we
derive the GRB X-ray flare data from {\em Swift/XRT}, and present
the fitting results. In section 3, we study some correlations
between parameters of X-ray flares. The distributions of GRB X-ray
flares and solar flares are discussed in Section 4. Conclusions and
discussion are given in Section 5. A concordance cosmology with
parameters $H_0 = 71$ km s $^{-1}$ Mpc$^{-1}$, $\Omega_M=0.30$, and
$\Omega_{\Lambda}=0.70$ is adopted in all part of this work.

\section{Data Analysis}

Since X-ray flares could be happened at any time of the afterglow
phase, the X-ray light curves usually contains one or more power-law
segments along with some flares. The mix of different components
makes the diverse X-ray afterglow light curves. Here we mainly focus
on the flare emission. We extensively search for the remarkable
feature of pulses at the GRB X-ray afterglow phase. We consider all
the {\em Swift} GRBs observed between 2005 April and 2015 March, and
select 199 GRBs during this period. These X-ray flares generally
contain a complete structure, including the remarkable rising and
decaying phase. These flares are clearly distinguishable from the
underlying continuum emission. We will also apply empirical
functions to fit the flare and the underlying component. Small
fluctuations around flare have not been identified as flares. The
total number of bright X-ray flares is 468, including 200 flares
with redshifts. Most GRBs contain a single or several flares. But
some of them have more than ten flares, such as GRBs 100212A and
100728A.

The 0.3-10 keV X-ray light curves of GRBs are taken from the website
of {\em Swift/XRT} (Evans et al. 2007, 2009)
\footnote{http://www.swift.ac.uk/xrt\_curves/}. We fit the flare
with a smooth broken power-law function (Li et al. 2012)
\begin{equation}\label{SBPL}
F_{1}(t)=F_{01}\left[\left(\frac{t}{t_{b}}\right)^{\alpha_{1}\omega}+\left(\frac{t}{t_{b}}\right)^
{\alpha_{2}\omega}\right]^{-\frac{1}{\omega}},
\end{equation}
and fit the underlying continuum with a power-law function (or broken power-law function)
\begin{equation}
F_{2}(t)=F_{02}t^{-\alpha_3},
\end{equation}
where $\alpha_{1}$, $\alpha_{2}$ and $\alpha_{3}$ are the temporal
slopes, $t_{b}$ is the break time, and $\omega$ represents the
sharpness of the peak of the light curve component. This method is
very similar to the fitting method of Chincarini et al. (2007,
2010). The two examples of the best-fitting flares are shown in
Figure 1. From this figure, we can see that the flares of GRB
060111A and GRB 080320 are well fitted. The fitting parameters of
flares, such as the start time, peak time, end time, peak flux,
fluence, peak luminosity and isotropic energy, are shown in Table 1.

Table 1 consists of 468 bright X-ray flares, including 200 flares
with redshifts. The time parameters of flares are derived as
follows. The rise time can be obtained by
$T_{rise}=T_{peak}-T_{start}$, the decay time
$T_{decay}=T_{end}-T_{peak}$ and the duration time
$T_{Duration}=T_{end}-T_{start}$, where $T_{start}$, $T_{peak}$ and
$T_{end}$ are the start time, peak time and end time of flares,
respectively. {The $T_{start}$ and $T_{end}$ are derived from
fitting temporal power-law curves to the rise and decay portions of
the flares. The points on the light curve where these power laws
intersect the underlying decay curve power law are defined as
$T_{start}$ and $T_{end}$. The definition is the same as Falcone et
al. (2007). However, similar as the definition of duration of prompt
emission, the time interval during which the integrated counts of a
burst go from 5\% to 95\% of the total integrated counts is more
reasonable. The waiting time is defined as
$T_{waiting}=T_{start,i+1}-T_{start,i}$, where $T_{start,i+1}$ is
the observed start time of the $i + 1th$ flare, and $T_{start,i}$ is
the observed start time of the $ith$ flare. All the flare properties
should be transferred into the source frame if they have redshift
measurements in the following analysis. For the first flare
appearing in X-ray afterglow, the rest-frame waiting time is taken
as $T_{start}/(1 + z)$. In the case of multiple flares, some flares
may occur during the activity of other flares. The waiting time can
also defined as above, because the start times of these flares are
different. This definition of waiting time is widely used in
geophysics (e.g., Omori 1895), magnetospheric physics (e.g., Chapman
et al. 1998), solar physics (e.g., Crosby 1996; Wheatland et al.
1998;), and astrophysics (e.g., Negoro et al. 1995; Wang \& Dai
2013). An extensive review on waiting time can be found in chapter 5
of Aschwanden (2011). The isotropic energy of flare in the 0.3-10
keV band can be obtain by $E_{x,iso}=4\pi D_{L}^{2} S_{F}/(1+z)$,
where $z$ is the redsihft, $D_L$ is the luminosity distance, and
$S_{F}$ is the fluence of flare. The flare fluence $S_F$ is
calculated by integrating the corresponding fitting smooth broken
power law function (equation (1)) from the start time to the end
time of the flare in the 0.2-10 keV energy band. The underlying
continuum has been subtracted. The peak luminosity and mean
luminosity can be derived through $L_{p}=4\pi D_{L}^{2} F_{p}/(1+z)$
and $L_{x,iso}=(1+z)E_{x,iso}/(T_{end}-T_{start})$, where $F_p$ is
the peak flux of flare.

\section{Parameters of X-ray Flares and Their Correlations}

Figure 2 shows the histogram distributions of the flare parameters.
The peak times of flares range from between 100 s and $10^6$ s after
GRB trigger, mainly from 100 s to 1000 s, at the early time of
afterglow phase. While according to Liang et al. (2010) and Yi et
al. (2013), the peak time of the optical onset bump is also in the
range of 100 - 1000 s statistically. Therefore the peak times of
flares are nearly matching the peak times of optical afterglow onset
bumps. The distributions of rise times and decay times are more
symmetric. Both of them are in the range 10 to $10^6$ s. The
isotropic energy of the X-ray flare with redshift can be estimated
from the fluence. The energy of flares mainly distributed from
$10^{50}$ erg to $10^{52}$ erg, about less than two orders of
magnitude compared with GRBs prompt emission. If the Gaussian
function is used to fit the 0.3-10 keV isotropic energy of X-ray
flares, the peak of the distribution is $10^{51.2}$ erg. Although
the distribution is quite skewed and the peak of the fitting does
not coincide with the peak of the energy distribution. The peak
luminosity of X-ray flares mainly range from $10^{48}$ erg $s^{-1}$
to $10^{50}$ erg $s^{-1}$, generally two or three orders of
magnitude larger than the peak luminosity of the optical afterglow
bumps. In the next paragraph, we will discuss the possible
correlations among those parameters of X-ray flares.

Figure 3 demonstrates the existence of a strong correlation
between the rise and decay times (the left one), i.e., $T_{decay}
\varpropto T_{rise}$, with the slope index 0.93. Generally, the
decay time is longer than the rise time, which is the general
property of shocks. There is also a strong correlation between the
flare duration time and the peak time. The duration times of X-ray
flares range from 10 s to $10^6$ s, and mainly distribute between
100 s and 1000 s. These two tight correlations suggest that longer
rise times associate with longer decay times, and also indicate
broader flares peak at later times.

We show some correlations among the characteristics of the
X-ray flares in the following figures. The correlations and linear
correlation coefficients from the Spearman pair correlation analysis
are shown in Table 2. Figure 4 shows correlations between the peak
luminosity and the flare time scales. These correlations clearly
demonstrate that a dimmer pulse of X-ray flare tends to peak at a
later time with a longer duration time. The correlation between peak
luminosity and the isotropic energy (the mean luminosity) indicates
that a flare with larger $E_{x,iso}$ ($L_{x,iso}$) tends to have a
brighter X-ray flare peaking at earlier time.

Figure 5 shows correlations between the mean luminosity and
the timescales of flares, which are transferred into the rest frame.
We obtain the mean luminosity thought the isotropic energy divided
by duration,  i.e., $L_{x,iso}=(1+z)E_{x,iso}/(T_{end}-T_{start})$.
We find that the mean luminosity is also tightly anti-correlated
with the time-scales of X-ray flares. These correlations between
mean luminosity and the timescales indicate that a dimmer X-ray
flare peaking at a later time also with a longer duration time.
Figure 6 shows the correlations between the waiting time and other
parameters of flares. The waiting time is correlated with both the
peak time and the duration time, which means a longer waiting time
tends to peak at a later time with a longer duration time. Besides,
the waiting time is anti-correlated with both the peak luminosity
and the mean luminosity. From above discussions, we conclude that
these correlations are consistent with each other.

\section{The Frequency Distributions of Flare Parameters}

The erratic X-ray flares generally supposed to be produced by the
late activities of central engine, therefore X-ray flares may share
a similar physical mechanism as GRB prompt emission. A lot of work
has been done to investigate their physical origin. Wang \& Dai
(2013) studied the distributions of the energies, duration times and
waiting times of solar flares and GRB X-ray flares. They found both
of them have similar statistical distributions. Apart from GRBs,
Wang et al. (2015) discovered that X-ray flares from the black hole
systems share the similar statistical properties with solar flares,
including Swift J1644+57, M87 and Sgr A$^*$. Solar flares are driven
magnetic reconnection (Lu \& Hamilton 1991; Charbonneau et al. 2001;
Morales \& Charbonneau 2008; Aschwanden 2012). The power-law
distributions of X-ray flare parameters indicate that they may be
self-organized criticality (SOC) events (Bak et al. 1987, 1988)
driven by magnetic reconnection. This suggests that GRB jet contains
a significant fraction of Poynting flux. The ratio $\sigma$ between
the Poynting flux $F_P$ and baryonic flux $F_b$ is larger than
unity, i.e., $\sigma=F_P/F_b\geq 1$ (Zhang \& Yan 2011). Meanwhile
the GRB prompt emission is likely powered by dissipation of magnetic
field energy (Lei et al. 2013; Yi et al. 2015b; Jia et al. 2015).
Recently, Uhm \& Zhang (2015) also studied the steep decay phase
after the peak time of X-ray flare and found the decay slope is
steeper than the standard value $\alpha=2+\beta$, where $\alpha$ and
$\beta$ are the decay slope of light curve and the observed spectral
index, respectively. This standard value can be understood as
follows. For a conical jet of GRB with an opening angle $\theta_j$,
emission from the same radius $R$ but from different viewing
latitudes $\theta\, (\theta < \theta_j)$ would reach the observer at
different times, which is called curvature effect. If the emission
area keeps a constant Lorentz factor $\Gamma$, there exists a simple
relation $\alpha=2+\beta$ (i.e., Kumar \& Panaitescu 2000). In the
same situation, the flare decay properties demand that the emission
region is undergoing significant bulk acceleration. In the
following, we will study the distributions of flare parameters using
a large sample.

We present the differential distributions of solar hard X-ray flares
in Figure 7. Because the number of solar flares is very large, we
select 11595 solar flares observed by {\em RHESSI} (Aschwanden
2011). We consider distribution of energy, waiting time, duration
time and peak flux. The number of flares $N(E)dE$ with energy
between $E$ and $E + dE$ can be expressed by
\begin{equation}\label{SBPL}
N(E)dE\varpropto E^{-\alpha_E}\,dE\,\,\,\,E<E_{max},
\end{equation}
where $\alpha_E$ is the power-law index and $E_{max}$ is the cutoff
energy. With this equation, we obtain the cumulative energy
distribution
\begin{equation}
N(>E)=a+b[E^{1-\alpha_{E}}-E_{max}^{1-\alpha_{E}}],
\end{equation}
where $a$ and $b$ are two fitting parameters. The power-law slope is
$\alpha_E=1.65\pm0.02$ for the differential distribution of solar
flares (Aschwanden 2011). For the other three parameters of solar
flares, their differential distribution can be expressed as
$N(X)dX\varpropto X^{-\alpha_X}$, where $X$ is corresponding to a
parameter of solar flares. All distributions show power-law
behavior. The indices of the waiting time, duration time and peak
flux are $2.04\pm0.03$, $2.00\pm0.05$ and $1.77\pm0.02$,
respectively (Aschwanden 2011). These distributions support that
solar flares are SOC events driven by a magnetic reconnection
process occurring in the atmosphere of the Sun. Therefore, the
distributions of solar flares could be well understood within a
physical framework, i.e., SOC.

We consider two groups of X-ray flares. In the first case, we
consider all 468 X-ray flares. Since some of them have no detected
redshift, we don't make any correction of the their time parameters.
To get the differential distributions of these varieties, we
separate them into 20 bins in the equal logarithmic space and then
use power-law function to fit them. We apply the Markov Chain Monte
Carlo (MCMC) technique to obtain the best fitting parameters and
give the $95\%$ confidential region. Besides, we also consider the
cumulative distribution of the peak flux of X-ray flares in this
case. We also use the fit function
$N(>F)=a+b[F^{1-\alpha_{F}}-F_{max}^{1-\alpha_{F}}]$ and the MCMC
method to obtain the optimal parameter and give the $95\%$
confidential region. The fitting result of the distribution of these
time parameters are shown in Figure 8. In the second case, we only
consider those X-ray flares with detected redshifts, which makes our
sample only include 200 X-ray flares. In this case, we transfer the
time parameters into the rest frame with $t_{rest}=t_{obs}/(1+z)$.
Moreover, because all of the X-ray flares in this sub-sample have
detected redshifts, we can calculate the isotropic energy through
$E_{iso}=4\pi D_{L}^{2} S_{F}/(1+z)$. The energy distributions GRB
X-ray flares show a flat part at the low energy regime, which could
be due to incomplete sampling and some selection bias for large
energy flares (Cliver et al. 2012). Therefore, in order to avoid
this selection effect, we only select the distribution above the
break to be fitted. Then, just same as the total sample, we use MCMC
method to obtain the optimal fit parameters and the $95\%$
confidential regions for the distributions. The results are shown in
Figure 9.

From Figure 8 and Figure 9, we find that the differential
distributions of the time parameters,  can be well fitted with the
power-law function both for the total sample and the sub-sample with
detected redshifts. In Figure 8, the power-law indices of the peak
times, rise times, decay times, waiting times and duration times for
the total sample are $-1.95 \pm 0.09$, $-1.56 \pm 0.04$, $-1.51 \pm
0.06$, $-1.89 \pm 0.13$, and $-1.56\pm 0.05$, respectively. While in
Figure 9, the power-law indices of the peak times, rise times, decay
times, waiting times and duration times for the sub-sample with
redshifts are $-1.72 \pm 0.11$, $-1.41 \pm 0.04$, $-1.33 \pm 0.08$,
$-1.44 \pm 0.05$, and $-1.41\pm 0.05$, respectively. There are a
little differences between the best-fitting parameters in the two
cases. For the cumulative distribution of peak flux and isotropic
energy of X-ray flares, we also obtain the power-law slopes. For the
total sample, we get the optimal parameter $\alpha_F=1.52\pm0.03$
for peak flux cumulative distribution. Meanwhile, for the
sub-sample, we get the optimal parameter $\alpha_E=1.32\pm0.07$ for
isotropic energy cumulative distribution. These power-law
distributions are natural predications of SOC theory (Aschwanden
2011).

Figure 10 shows the power-law distributions of solar flares and GRB
X-ray flares. Although the energies of GRB X-ray flares are in the
range from $10^{49}$ erg to $10^{53}$ erg, and $10^{28}$ erg to
$10^{32}$ erg for solar flare energies, they show power-law
distributions with different indices. These distributions can be
understood in fractal-diffusive avalanche model (Aschwanden 2012).
The magnetic reconnection as physical origin of solar flares is well
recognized (Sweet 1958; Parker 1957; for a recent review, see
Shibata \& Magara 2011). For solar flares, the total magnetic energy
$E_B$ released during a reconnection process in an elementary volume
$L^3$ with an average magnetic energy density $B^2/8\pi$ is (Shibata
\& Magara 2011)
\begin{equation}
E_B=L^3\frac{B^2}{8\pi}\sim3\times10^{30}(\frac{B}{10^2G})^2
(\frac{L}{2\times10^9 cm})^3~~ \rm erg,
\end{equation}
which is the typical energy of solar flare. For models of
GRB X-ray flares, there are several magnetic reconnection models.
Dai et al. (2006) proposed that X-ray flares could be produced by
differentially rotating, millisecond pulsars. The differential
rotation leads to windup of interior poloidal magnetic fields and
the toroidal fields break through the stellar surface. The energy
from reconnection toroidal fields with different polarity is (Dai et
al. 2006; Kluzniak \& Ruderman 1998)
\begin{equation}
E_b=\frac{E_b^2}{8\pi}V_b\sim1.6\times10^{51}\frac{V_b}{V_*}~\rm
erg,
\end{equation}
where $V_b$ and $V_*$ are the toroid's volume and the
stellar volume, respectively. This energy is comparable to the
observed one. X-ray flares of GRBs may also be powered by magnetic
dissipation in a decelerating shell (Giannios 2006). The analysis by
Giannios (2006) shows that the energy emitted in a single flare
$E_f$ and produced by a single reconnection event is
\begin{equation}
E_f\leq 5\epsilon\frac{E_{fs}}{\delta^2} \sim 1.25\times
10^{52}(\frac{E_{fs}}{10^{53}erg})(\frac{\delta}{2})^{-2}~\rm erg,
\end{equation}
where $E_{fs}$ is the isotropic energy of the forward shock,
and $\epsilon \sim0.1$ the fraction of the Alfv\'{e}n speed of the
magnetic reconnection in a strongly magnetized plasma. For a
constant-density medium, the typical value of $\delta$ is four
(Waxman 1997). In stellar wind case, the value of $\delta$ is two
(Pe'er \& Waxman 2005). So the energy budgets of these models are
comparable to the observed energies of X-ray flares.

\section{Conclusions and Discussion}
In this paper, we present a catalog of 468 bright X-ray flares of
GRBs taken from the online {\em Swift/XRT} GRB Catalogue until March
2015, including 200 flares with redshifts. We use a smooth broken
power-law function to fit the X-ray flares, and obtain the fitting
parameters of X-ray flares, which are listed in Table 1. The peak
times of flares range from between 100 s and $10^6$ s after GRB
trigger, mainly from 100 s to 1000 s. The duration times of flares
also mainly distribute between 100 s and 1000 s. The 0.3-10 keV
isotropic energy of X-ray flares mainly distributed from $10^{50}$
erg to $10^{52}$ erg, which is about less than two or three orders
of magnitude compared with GRBs prompt emission. The 0.3-10 keV
isotropic energy of X-ray flares distribution is a lognormal peaked
at $10^{51.2}$ erg.  We also found some tight correlations between
these parameters of X-ray flares, and the best-fitting results for
the correlations are shown in Table 2. Generally, these correlations
clearly demonstrate that a dimmer pulse of X-ray flare tends to peak
at a later time with a longer duration time.

We also study the frequency distributions of solar flares and GRB
X-ray flares. In the analysis, we combine all X-ray flares from long
and short GRBs. Some studies indicated that X-ray flares in long and
short GRBs may have a common origin (Margutti et al. 2011; Wang \&
Dai 2013). The best-fitting results for the power-law distributions
of these parameters are shown in Figures 7, 8, and 9. We find there
are four power-law distributions with different indices between
X-ray flares and solar flares, including power-law distributions of
energies, durations, peak fluxes and waiting times. These
distributions could be explained by a fractal-diffusive,
self-organized criticality model. Besides, we also investigate the
peak times, rising times and decay times of X-ray flares, and find
all of them show power-law distributions.

Interestingly, the ratio $T_{duration}/T_{peak}$ is almost constant
with time in our X-ray flare sample (see right panel of Fig. 3).
This result is also found by Chincarini et al. (2010). The late
internal shock model can not account for this result since the
arrival time is not related to the collision conditions (Kobayashi
et al. 1997; Ramirez-Ruiz \& Fenimore 2000). Further more, the
efficiency of internal collision is typically low (Panaitescu et al.
1999; Kumar 1999; Fan \& Wei 2005).

X-ray flares may also be powered by magnetic dissipation in a
decelerating shell (Giannios 2006). MHD instability could be
triggered in strongly magnetized ejecta during its deceleration due
to interaction with the external medium. This instability can
release energy through magnetic reconnection. Multiple flares are
expected because of dissipation in multiple neighboring regions in
the decelerating flow. This model also predicts that smooth flares
are more energetic than spiky ones (Giannios 2006). Chincarini et
al. (2010) compared this model with X-ray flare data, and found this
model is not in contradiction with observation. However more test
are needed. Another possible model for X-ray flares is the
internal-collision-induced magnetic reconnection and turbulence
model, which can also reproduce the properties of GRB prompt
emission (Zhang \& Yan 2011). In this model, internal collisions
distort the ordered magnetic field lines in the ejecta. The X-ray
flares can be triggered by magnetic reconnection in the distorted
magnetic field. So these two models are favored from our analysis.
However, much more data are required to constrain the model
parameters.

\section*{Acknowledgments}
We thank an anonymous referee for useful suggestions and comments.
We many thank Zi-Gao Dai and Bing Zhang for valuable comments. This work is
supported by the National Basic Research Program of China (973
Program, grant No. 2014CB845800) and the National Natural Science
Foundation of China (grants 11422325, 11373022, 11533003 and
11163001), the Excellent Youth Foundation of Jiangsu Province
(BK20140016).

\begin{figure*}
\includegraphics[angle=0,scale=0.31]{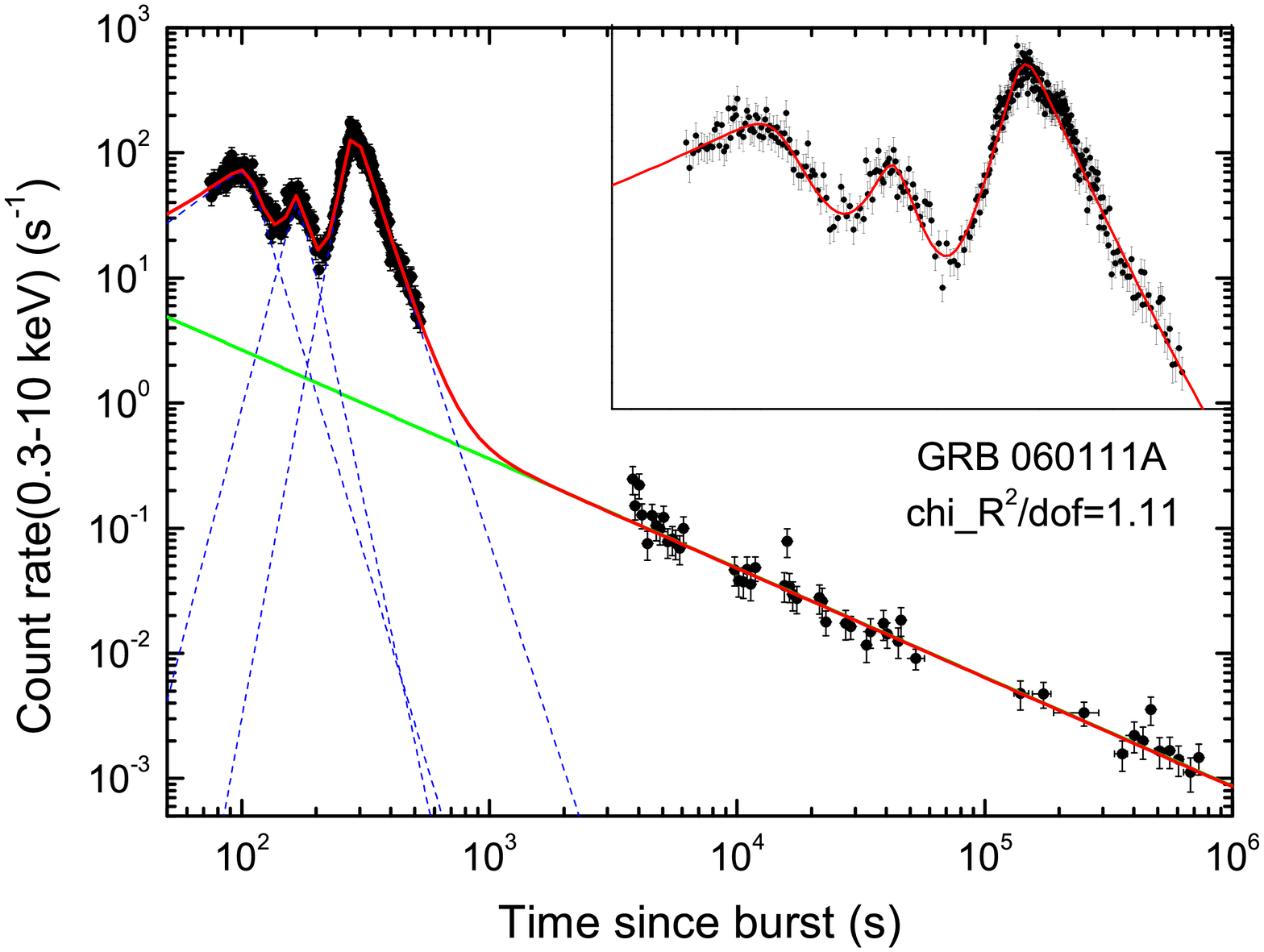}
\includegraphics[angle=0,scale=0.31]{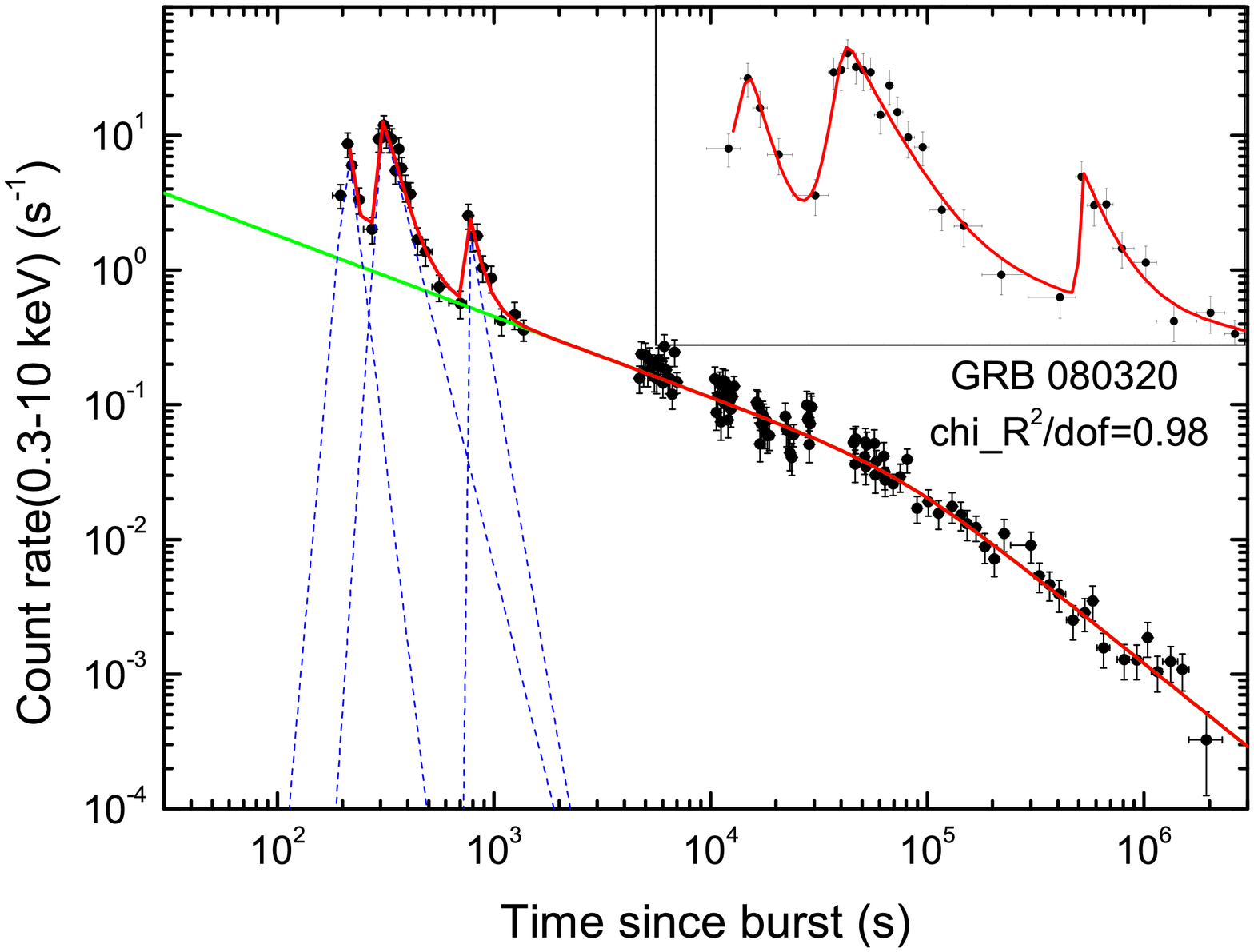}
\caption{Best fitting for the X-ray flares of GRB 060111A and GRB
080320. The blue-dash lines show the best fitting for individual
flares, and the green line shows the underlying continuum. The red
line shows the total best fitting. Inset: the detail of flare
fittings.}
\end{figure*}

\clearpage
\begin{figure*}
\includegraphics[angle=0,scale=0.30]{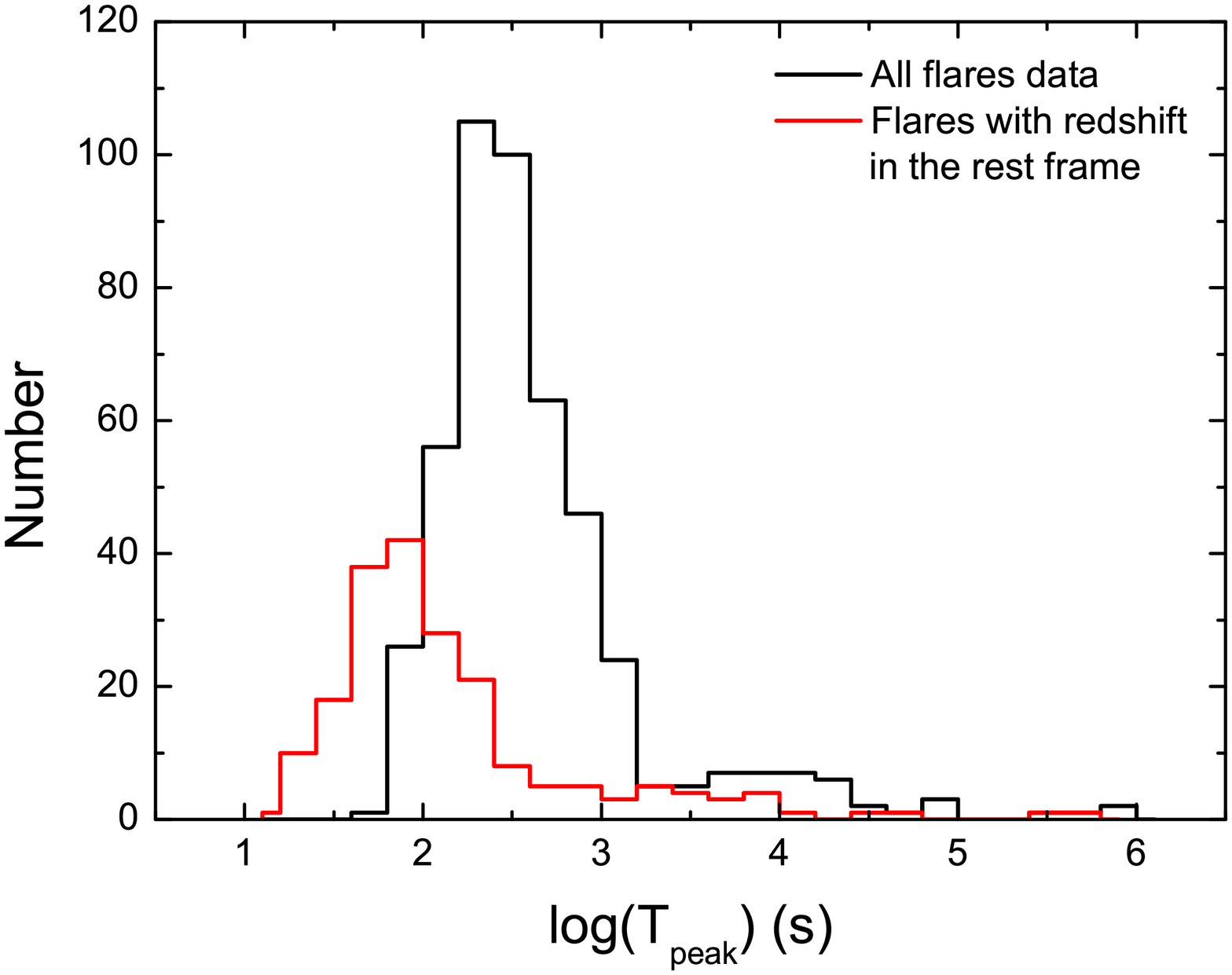}
\includegraphics[angle=0,scale=0.30]{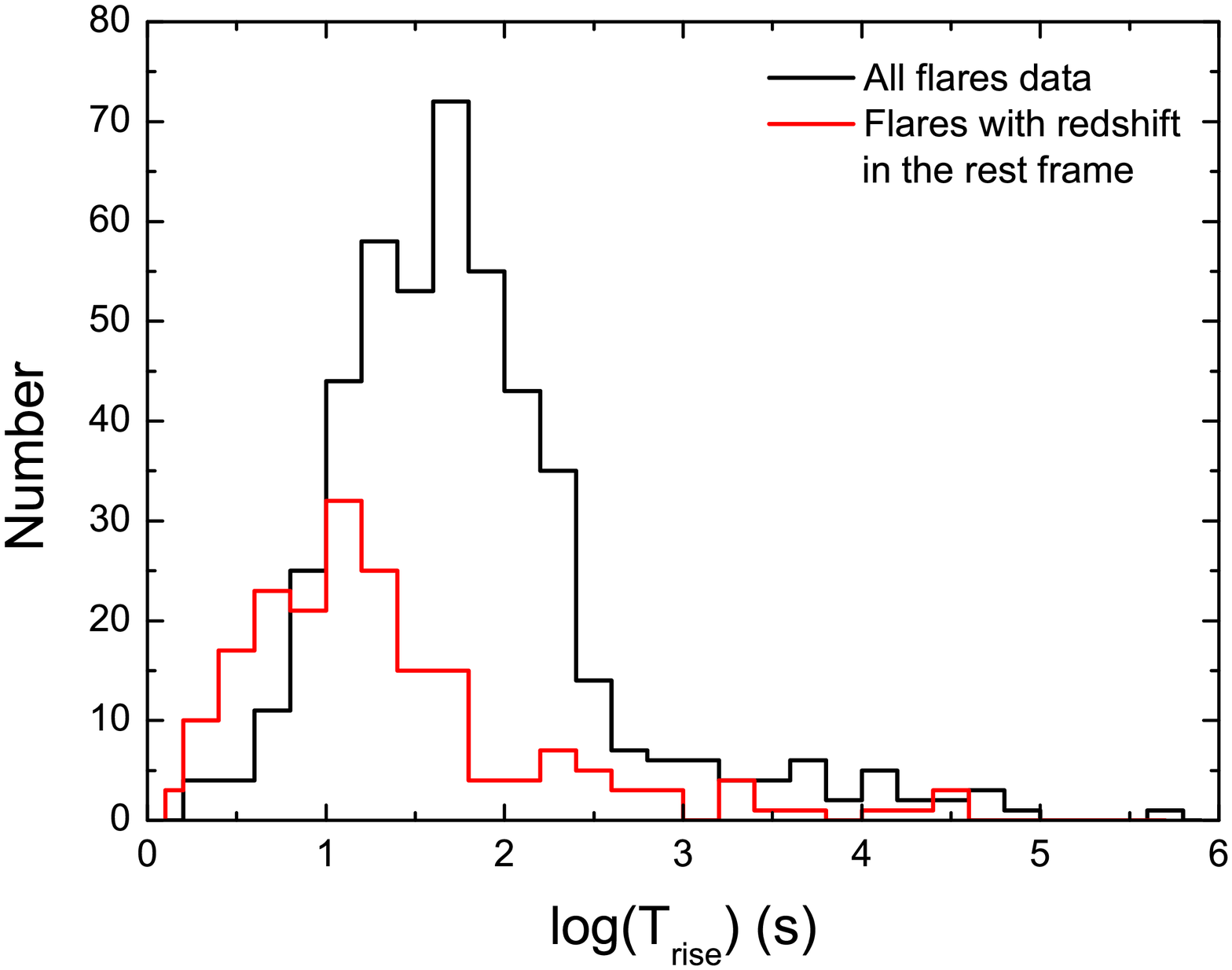}
\includegraphics[angle=0,scale=0.30]{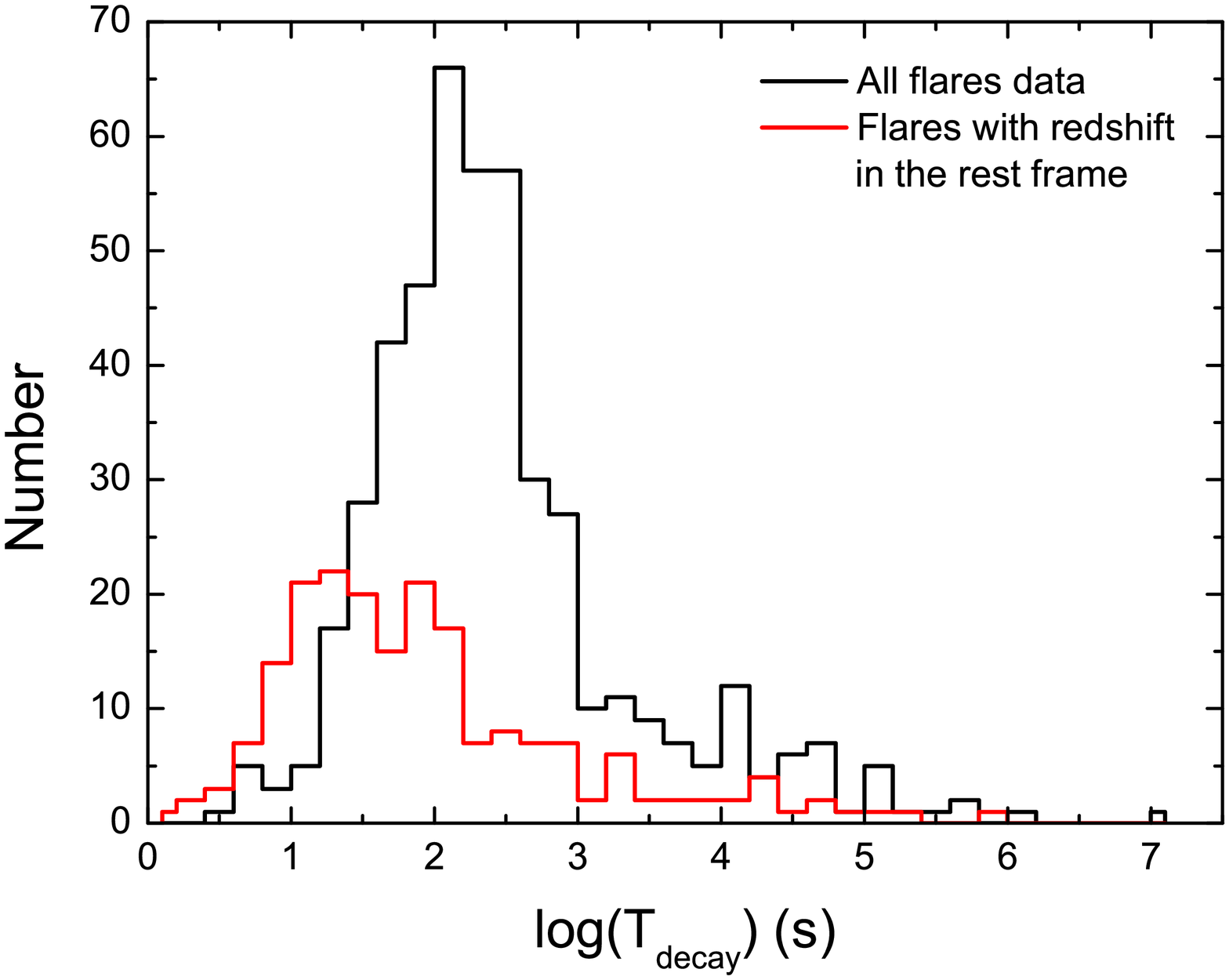}
\includegraphics[angle=0,scale=0.30]{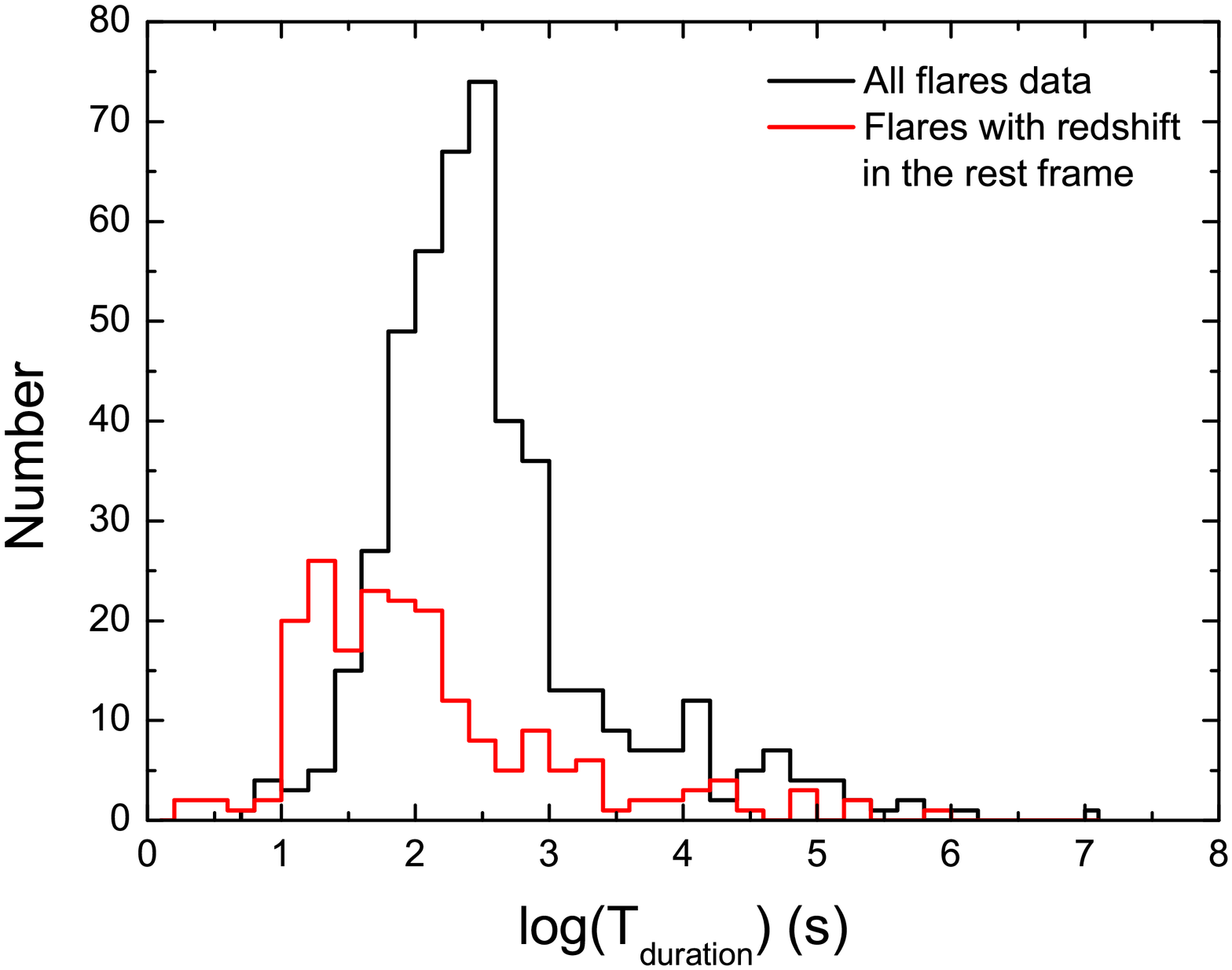}
\includegraphics[angle=0,scale=0.30]{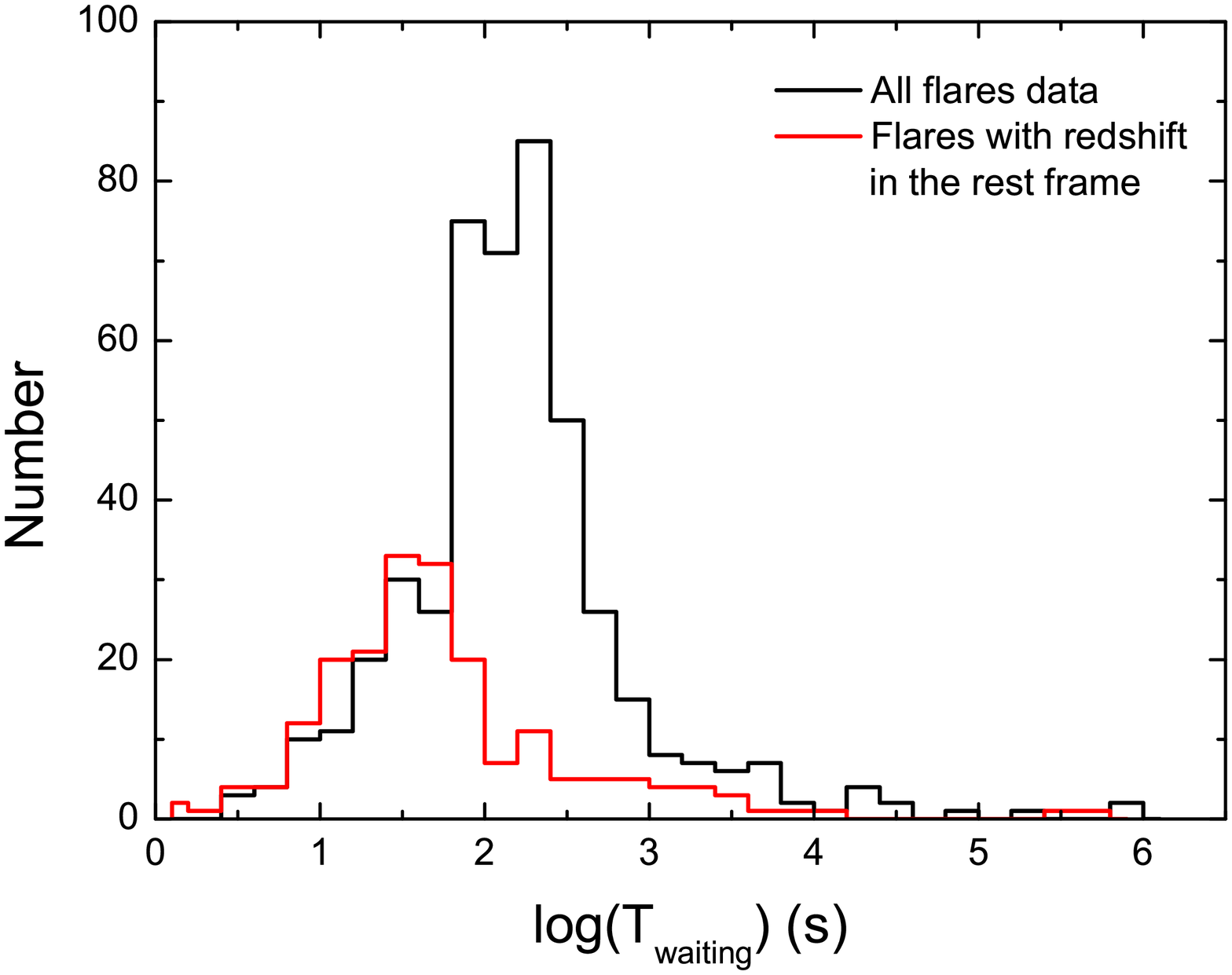}\hfill
\includegraphics[angle=0,scale=0.30]{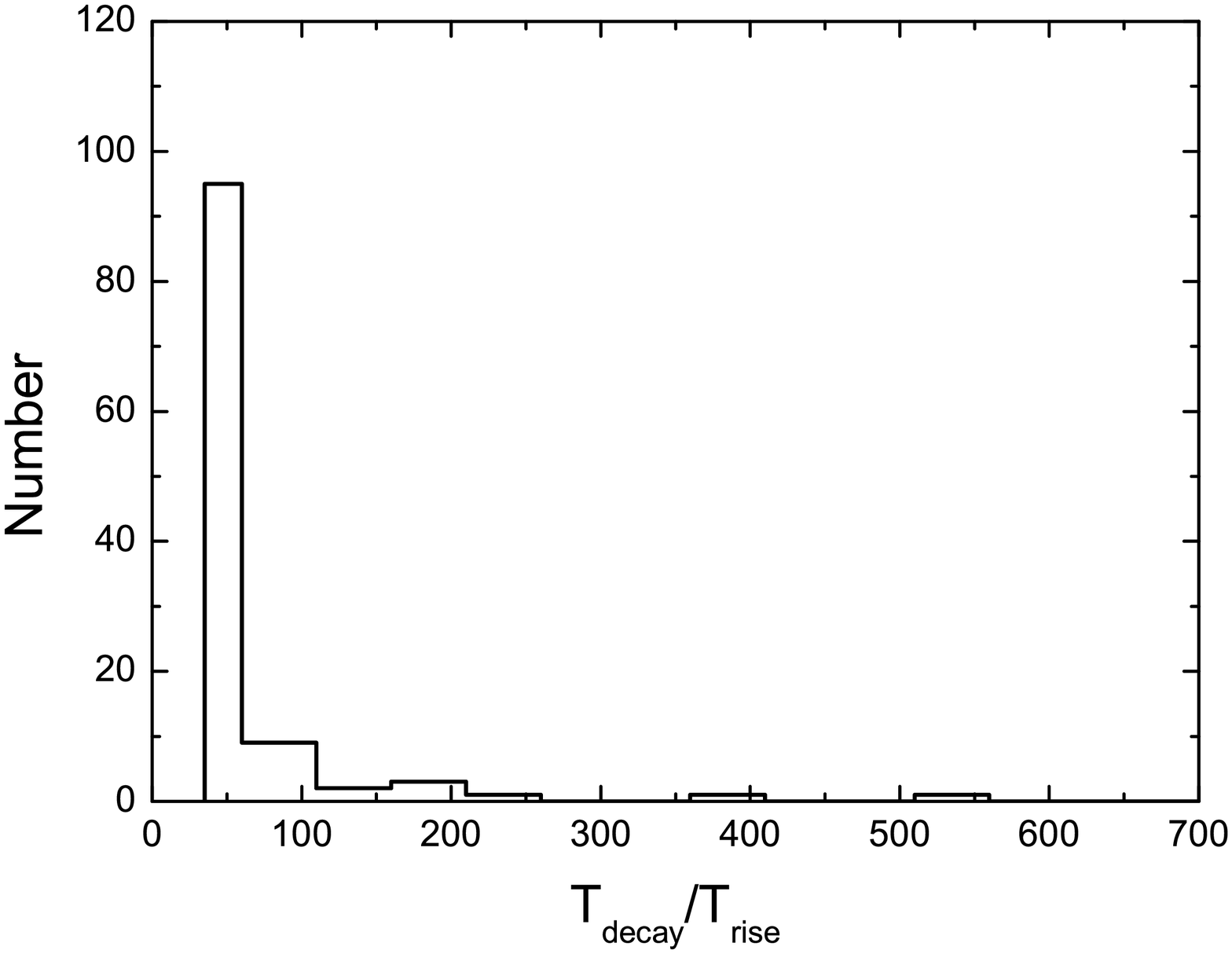}
\caption{The histogram distributions of X-ray flare parameters. For
the observed time parameters, the black line is corresponding to the
fitting results of all the flare sample. While the red line
represent the flares with reshifts, and the parameters have been
transferred to the source frame. The mean value is 24.64 for
$T_{decay}/T_{rise}$. The blue line shows the best lognormal fit of
energy.}
\end{figure*}

\begin{figure*}
\includegraphics[angle=0,scale=0.30]{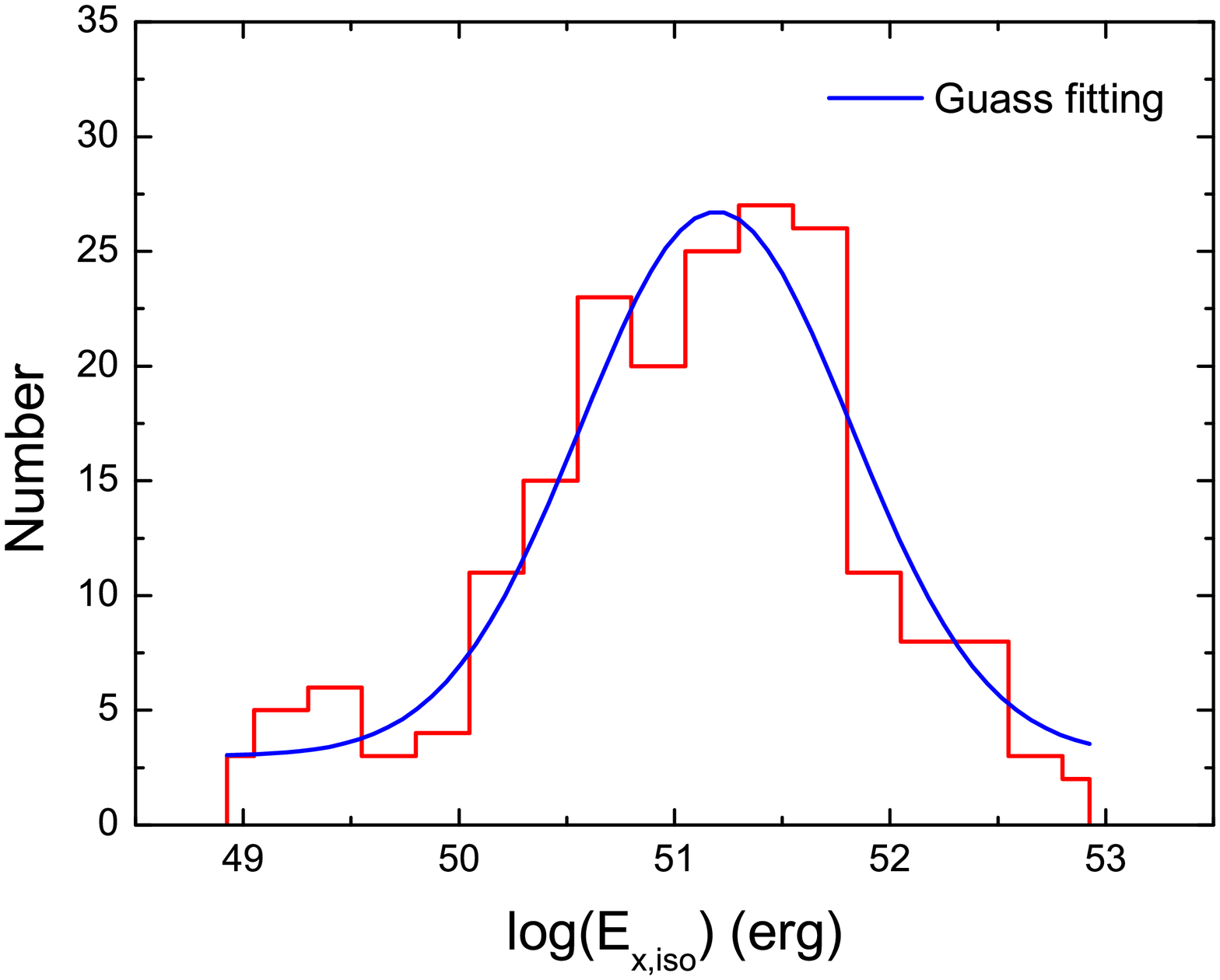}
\includegraphics[angle=0,scale=0.30]{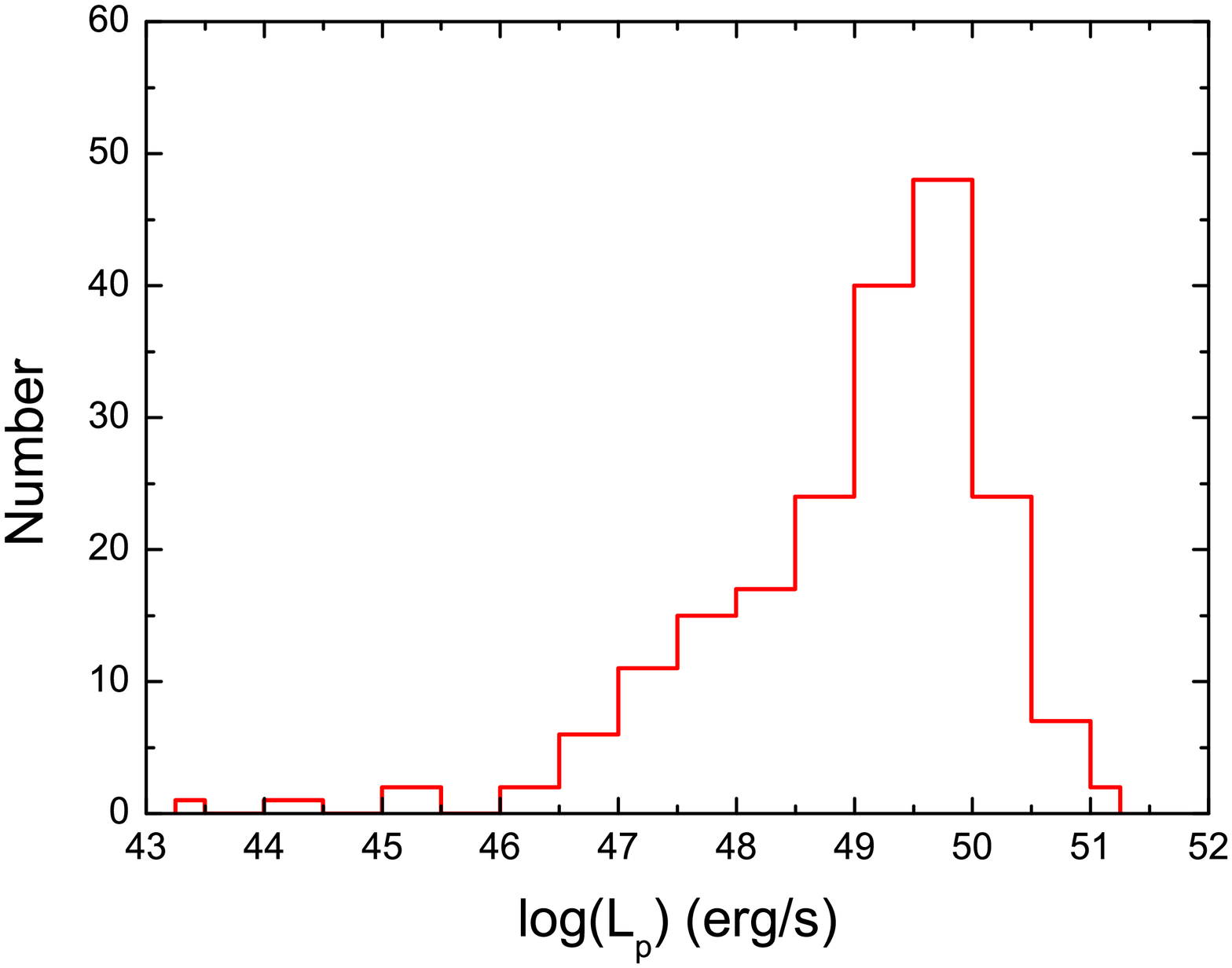}
\center{Fig. 2--- Continued}
\end{figure*}

\clearpage
\begin{figure*}
\includegraphics[angle=0,scale=0.30]{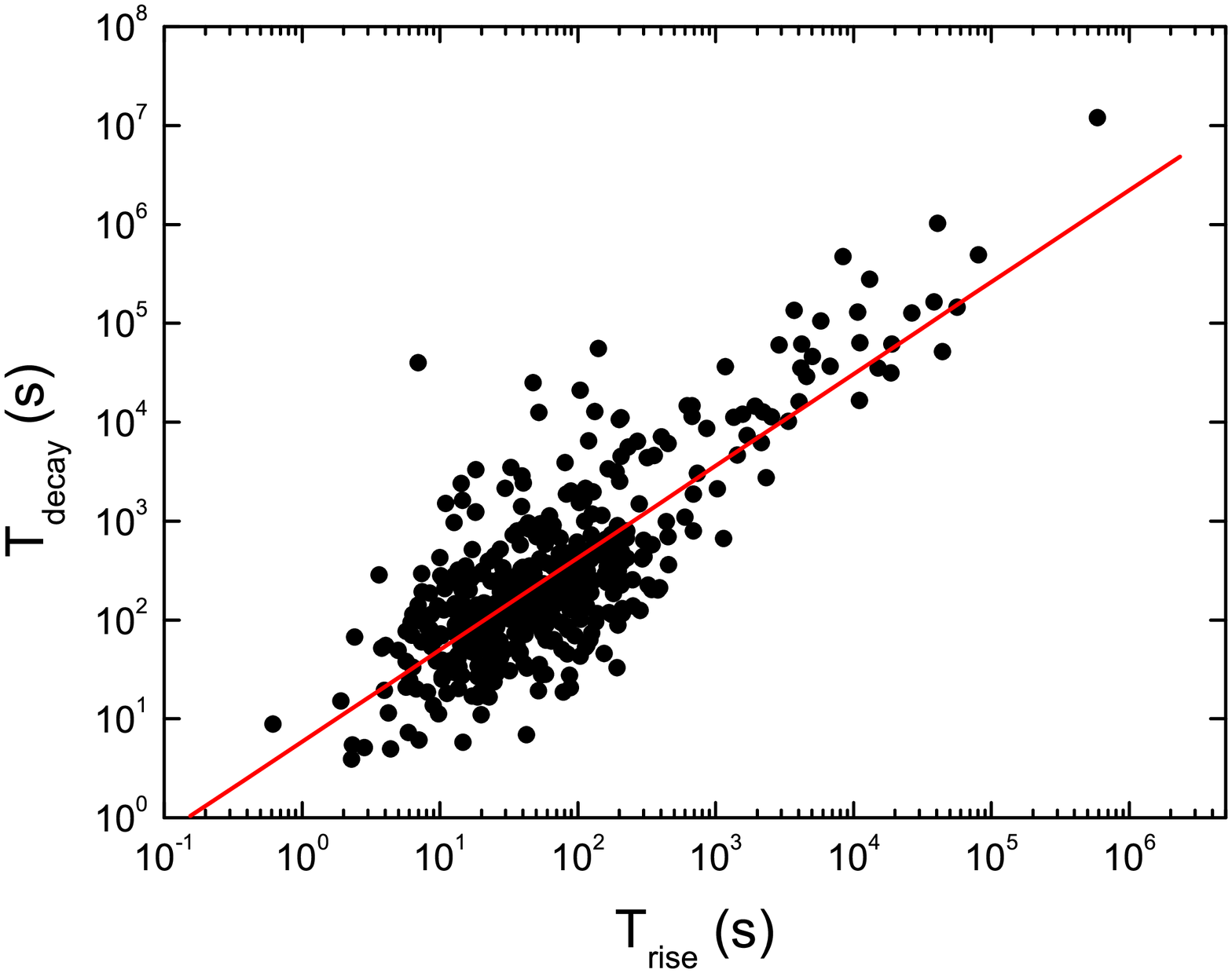}
\includegraphics[angle=0,scale=0.30]{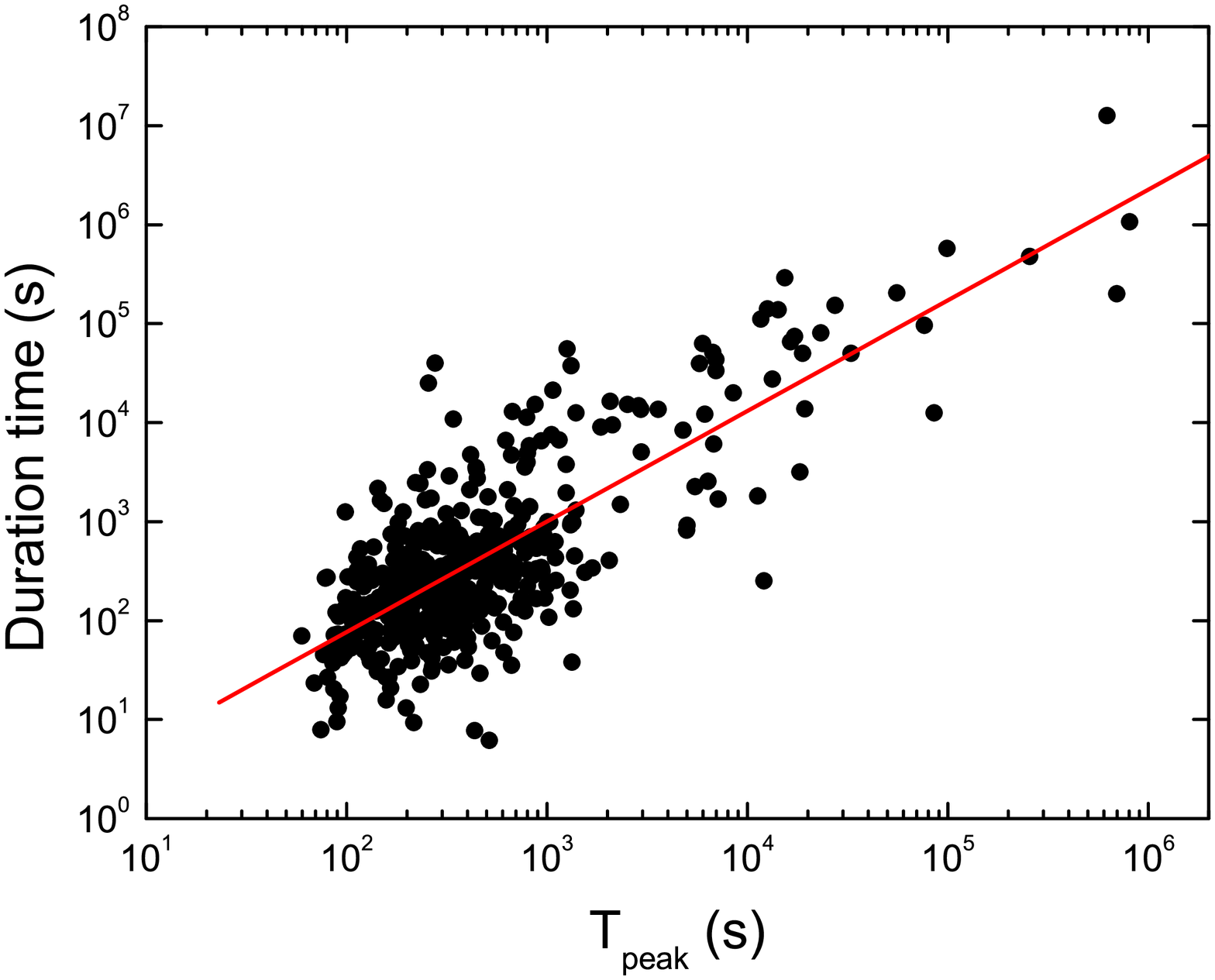}
\caption{The correlations of time-scales of GRB X-ray flares. Left
panel: the rise time is correlated with the decay time. Right panel: the duration is
correlated with peak time. The red line is the best fitting. The
best fitting results are shown in Table 2.}
\end{figure*}

\clearpage
\begin{figure*}
\includegraphics[angle=0,scale=0.30]{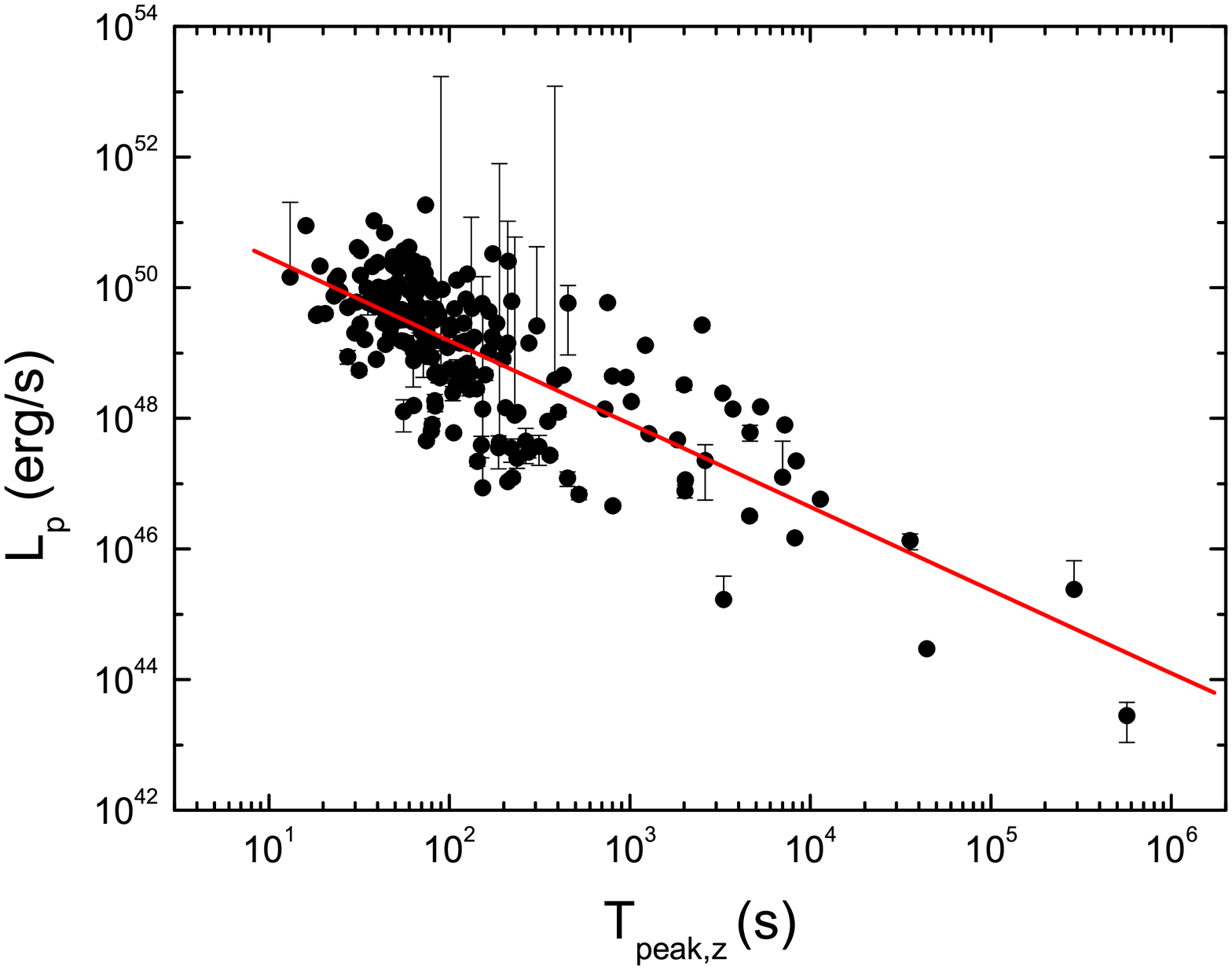}
\includegraphics[angle=0,scale=0.30]{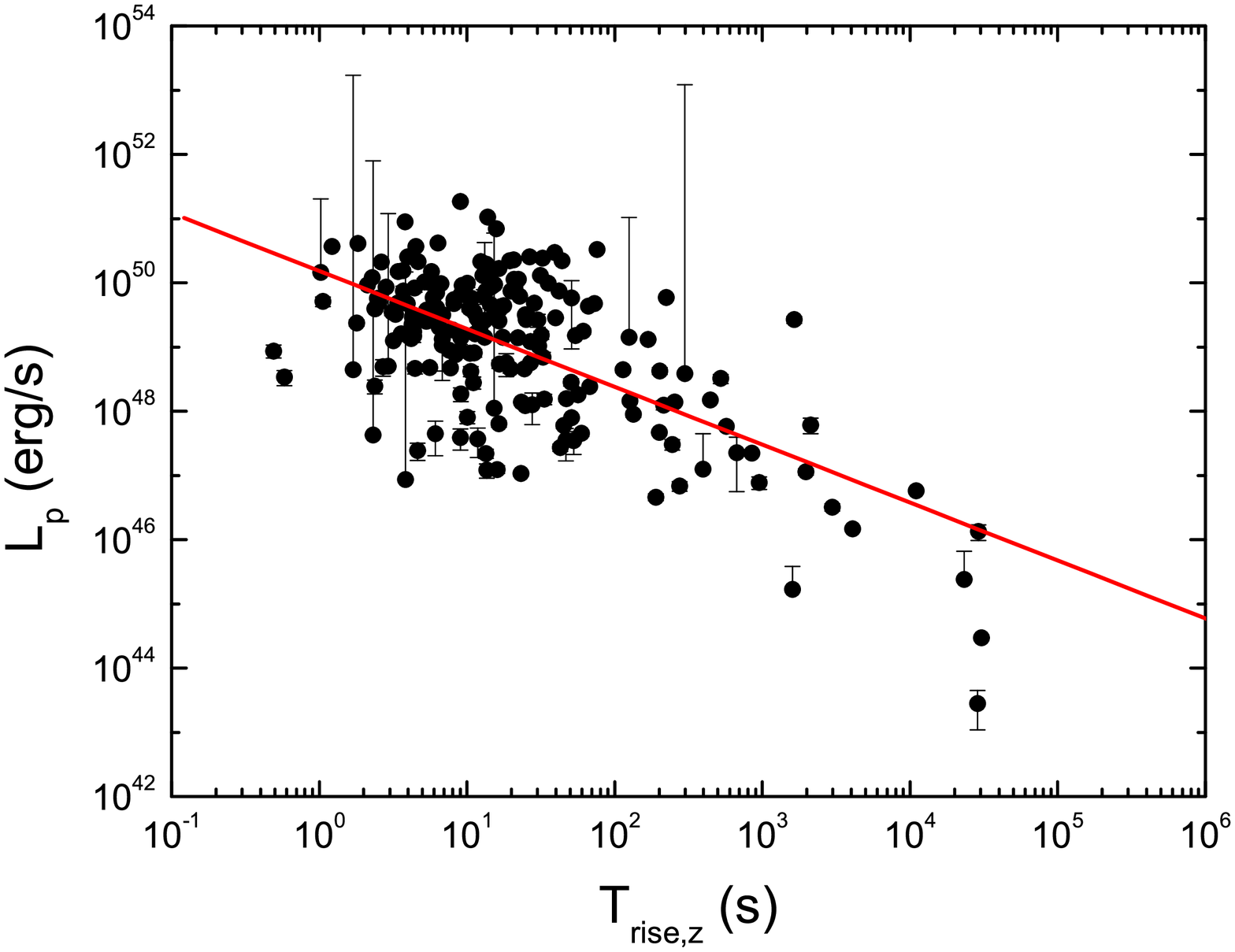}
\includegraphics[angle=0,scale=0.30]{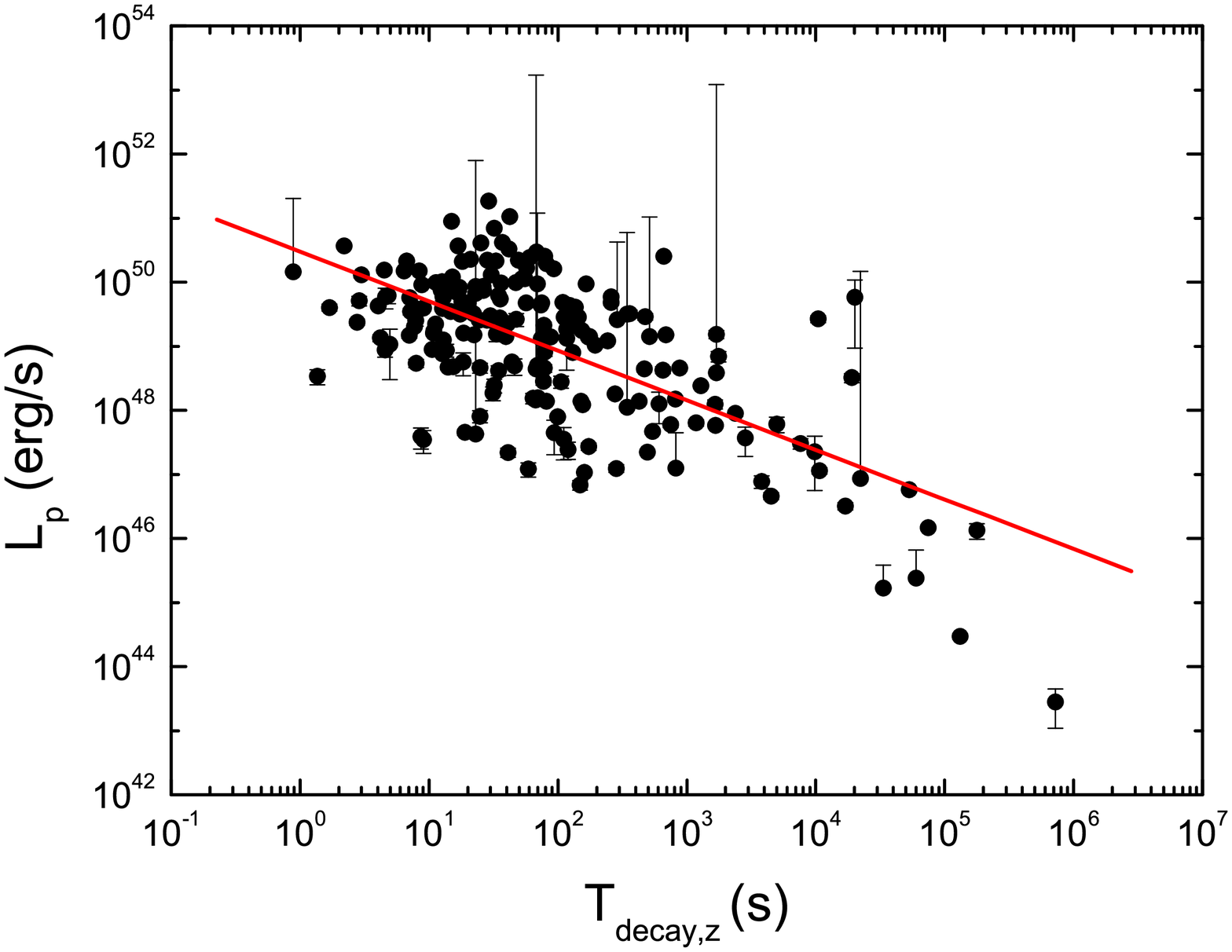}
\includegraphics[angle=0,scale=0.30]{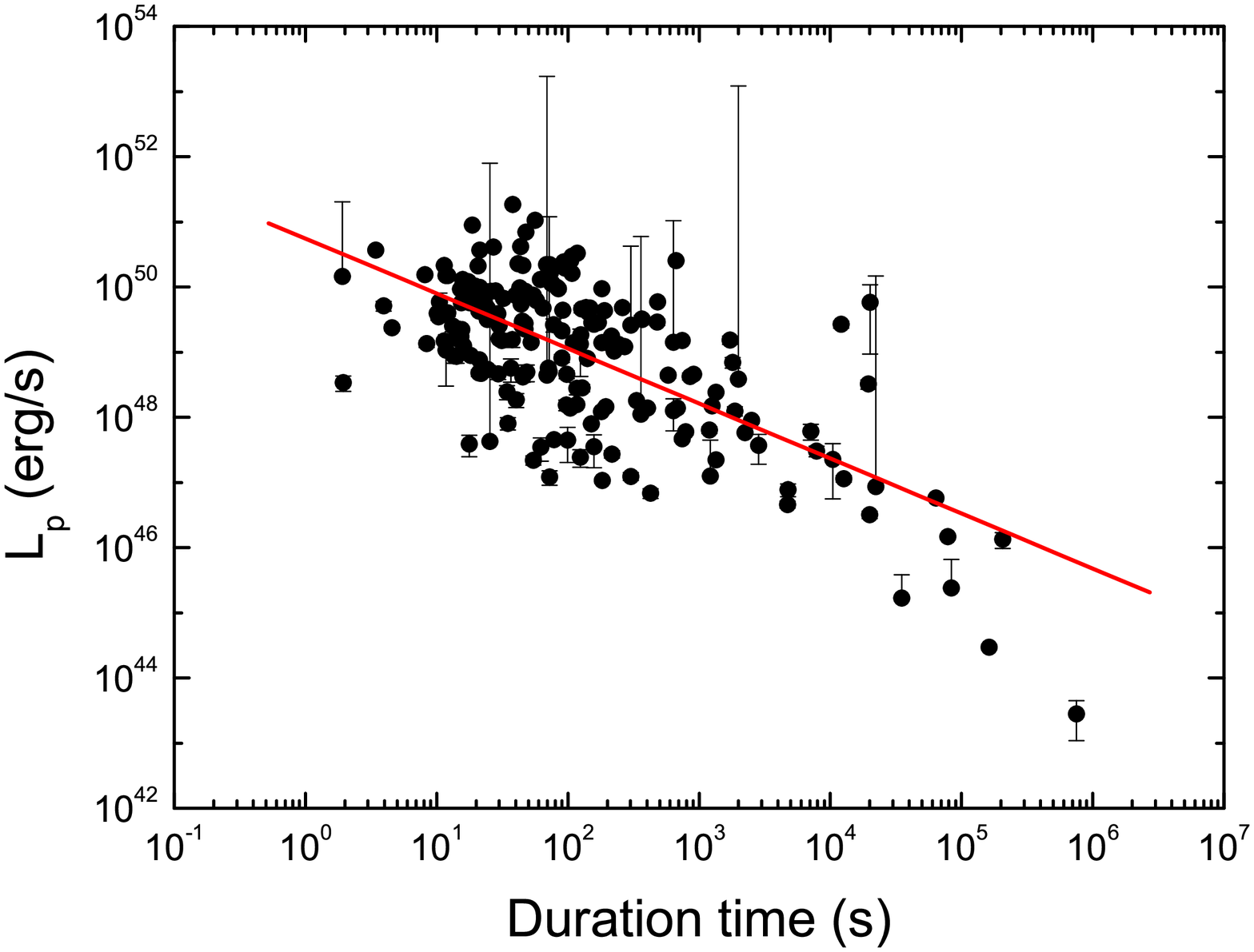}
\includegraphics[angle=0,scale=0.30]{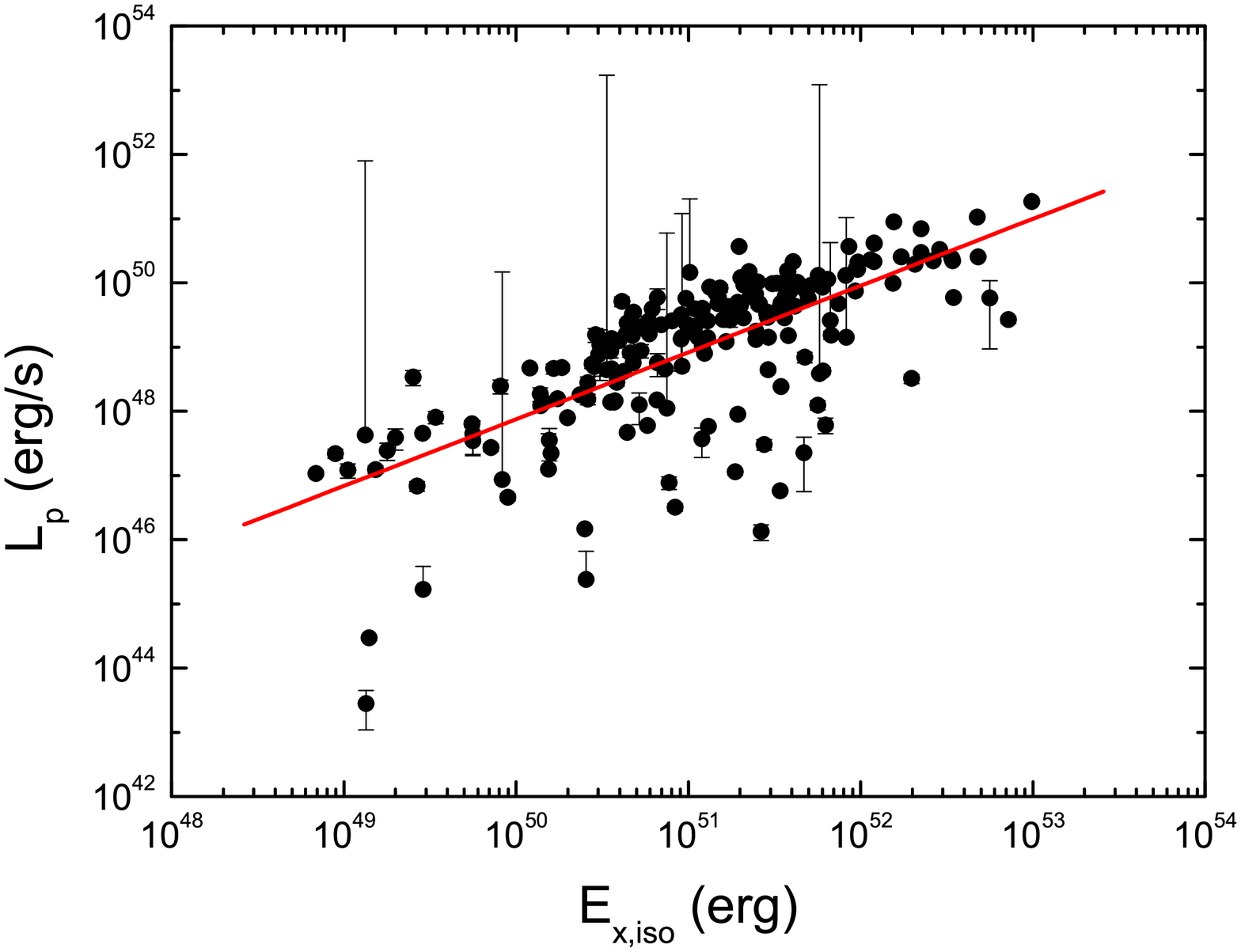}
\includegraphics[angle=0,scale=0.30]{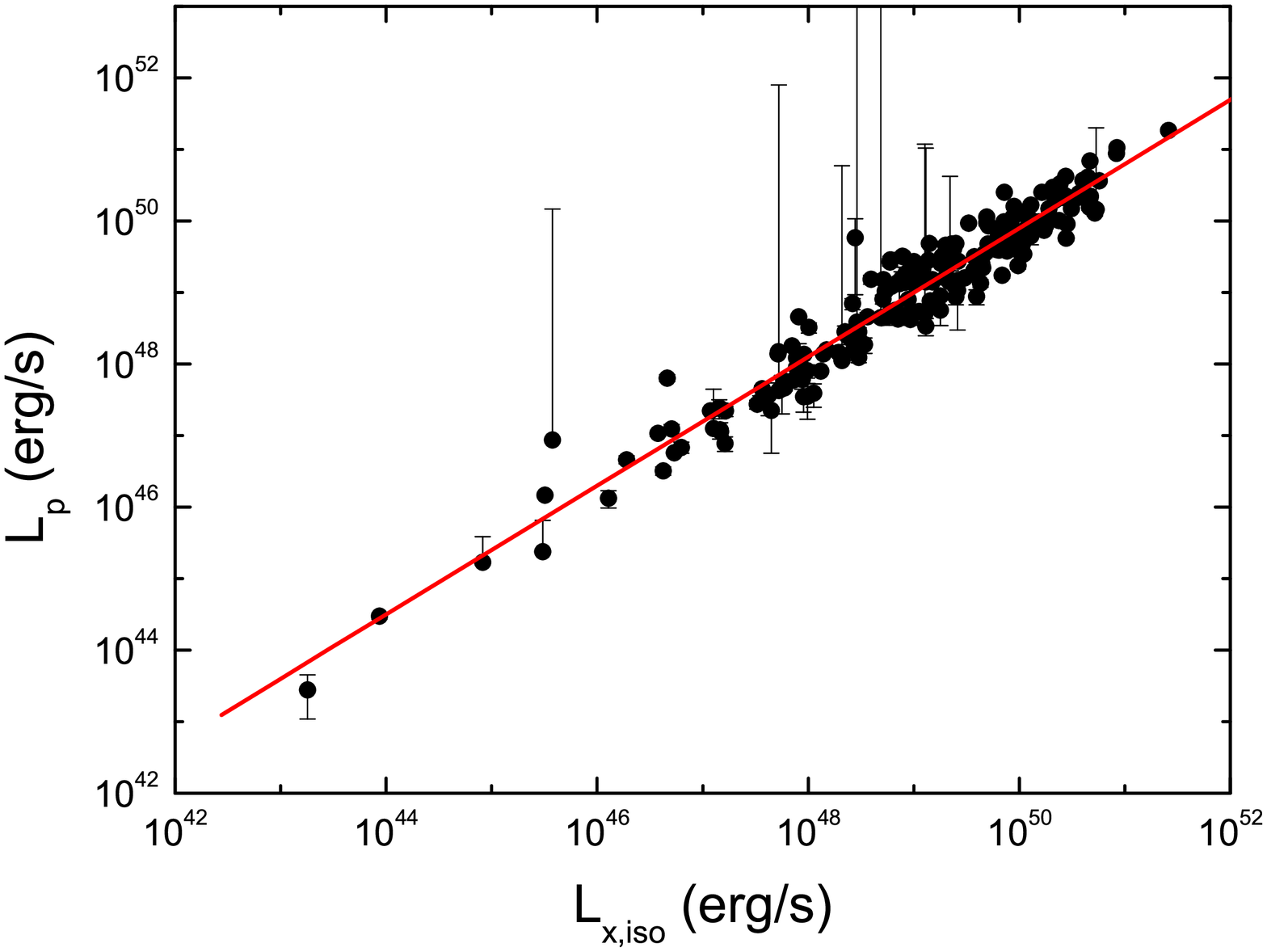}
\caption{The correlations between peak luminosity and other
parameters of GRB X-ray flares. The red line is the best fitting.
The times are transferred into the source frame. The best fitting
results are listed in Table 2.}
\end{figure*}

\clearpage
\begin{figure*}
\includegraphics[angle=0,scale=0.30]{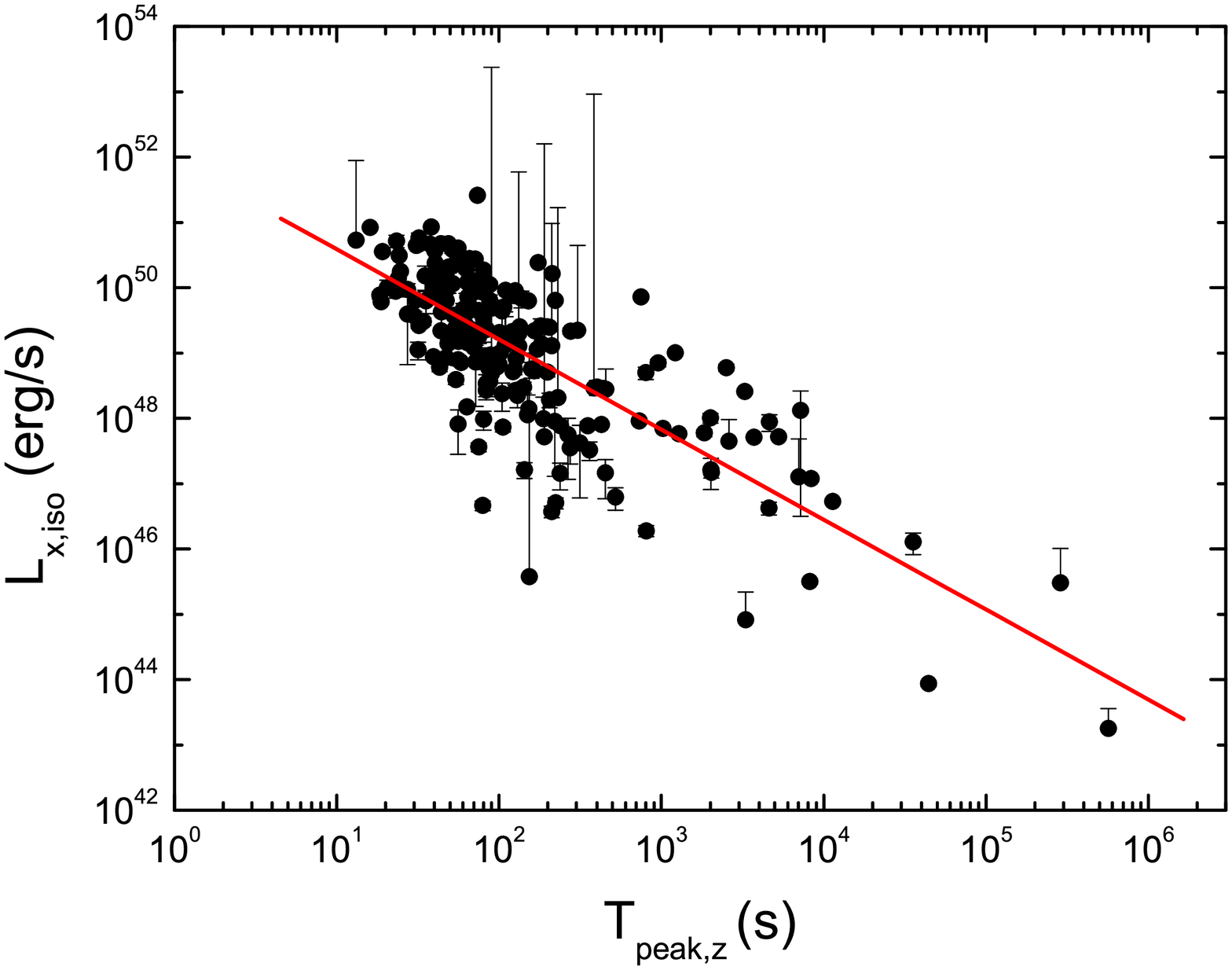}
\includegraphics[angle=0,scale=0.30]{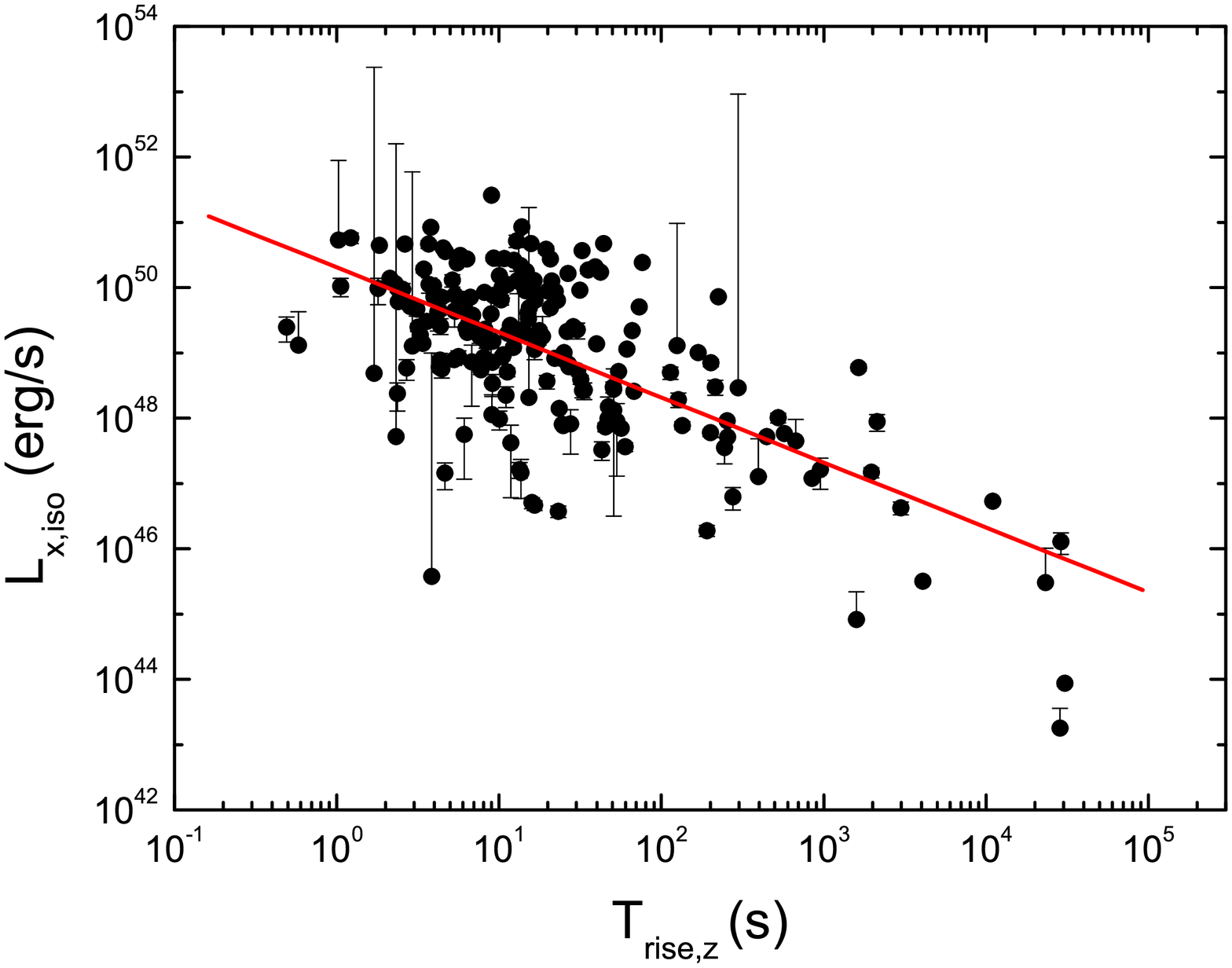}
\includegraphics[angle=0,scale=0.30]{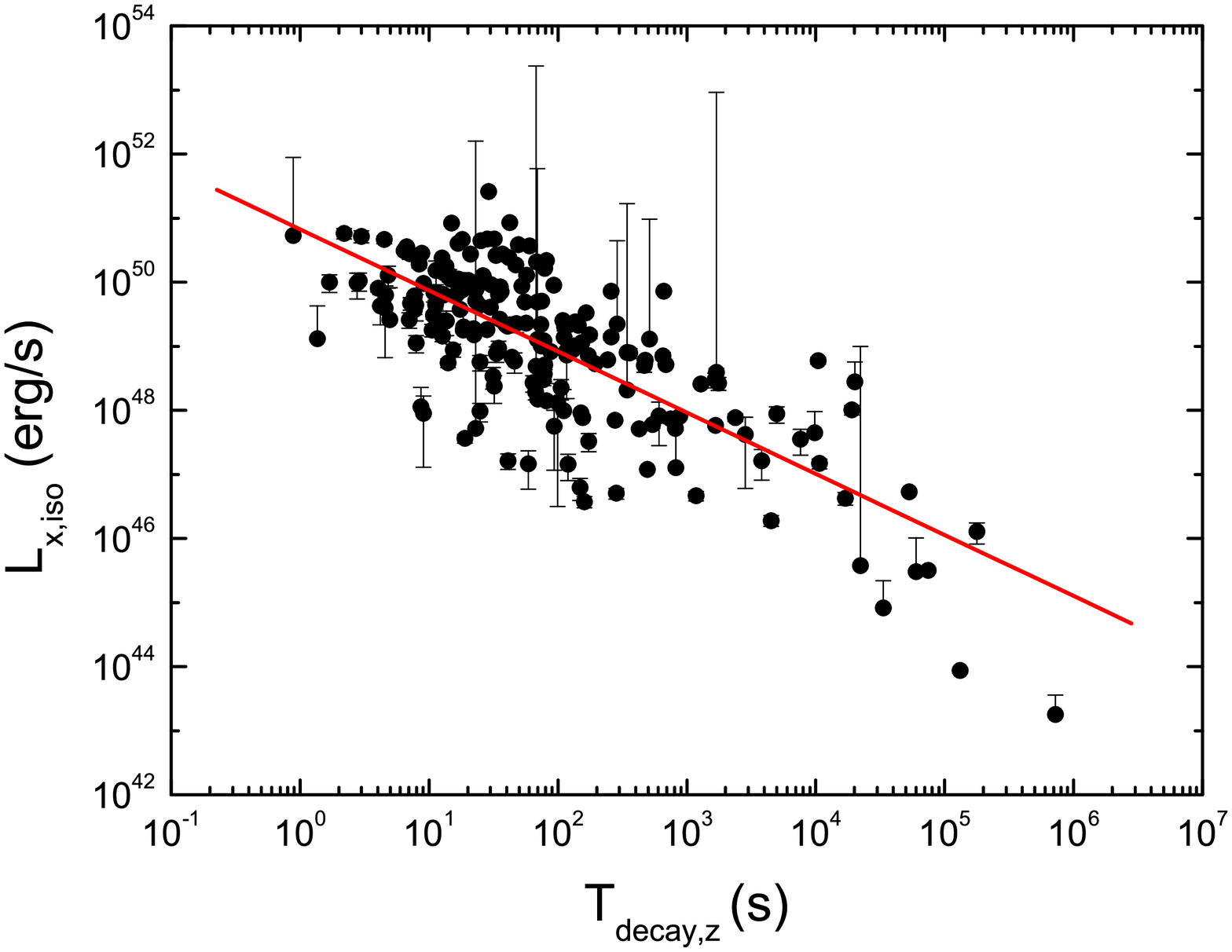}
\includegraphics[angle=0,scale=0.30]{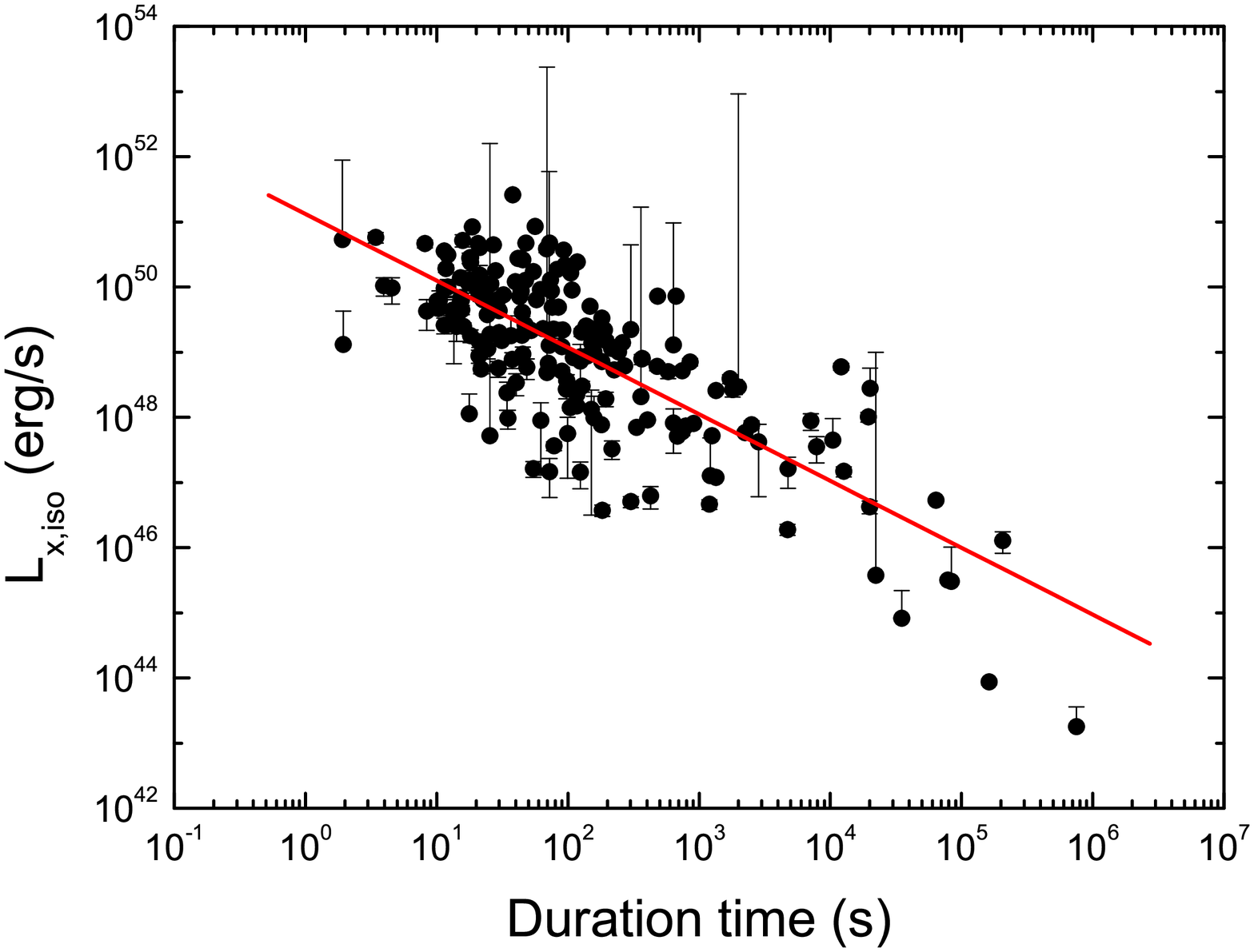}
\caption{The correlations between mean luminosity and time scales of
GRB X-ray flares. Red lines are the best fittings. The best fitting
results can be seen in Table 2.}
\end{figure*}

\clearpage
\begin{figure*}
\includegraphics[angle=0,scale=0.30]{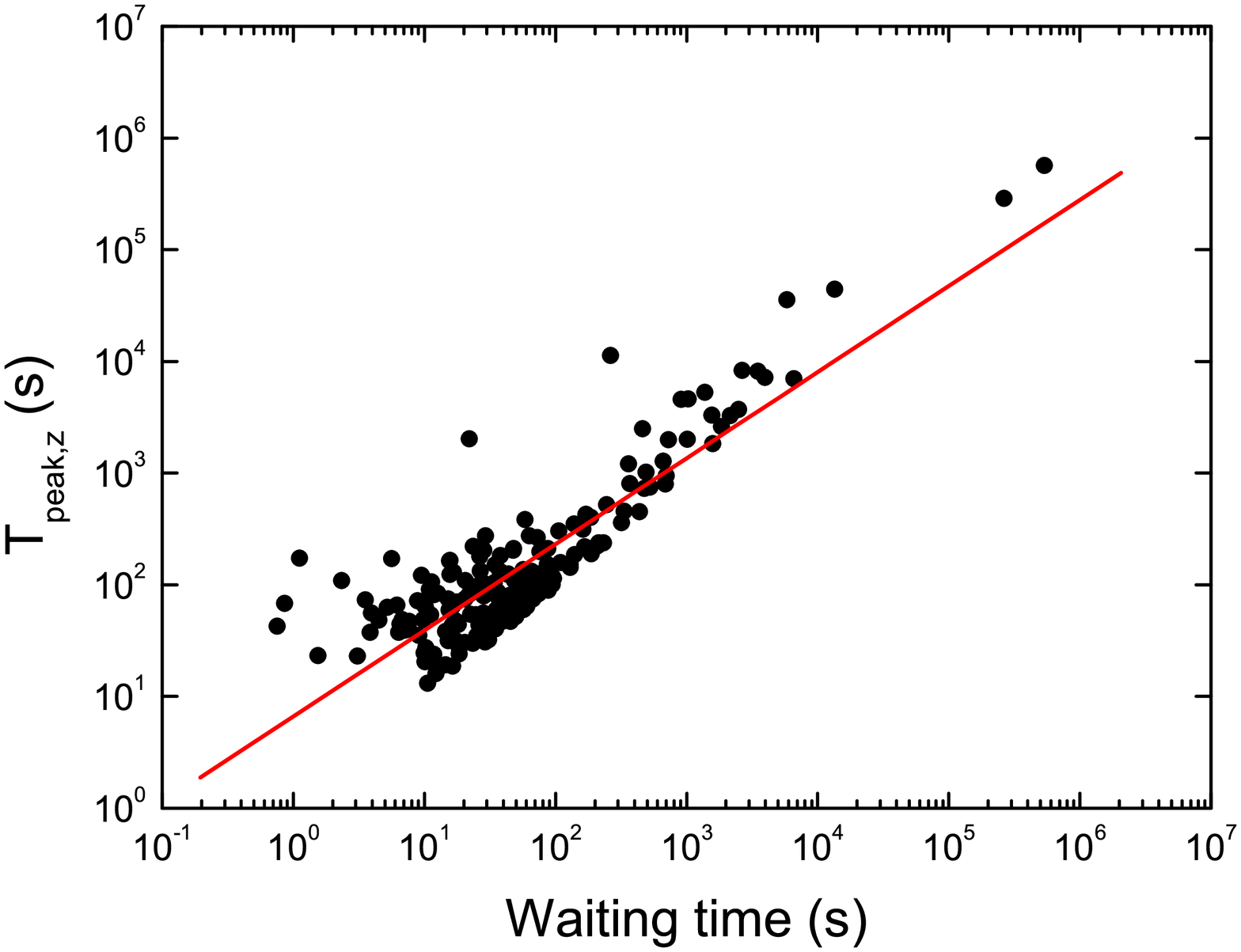}
\includegraphics[angle=0,scale=0.30]{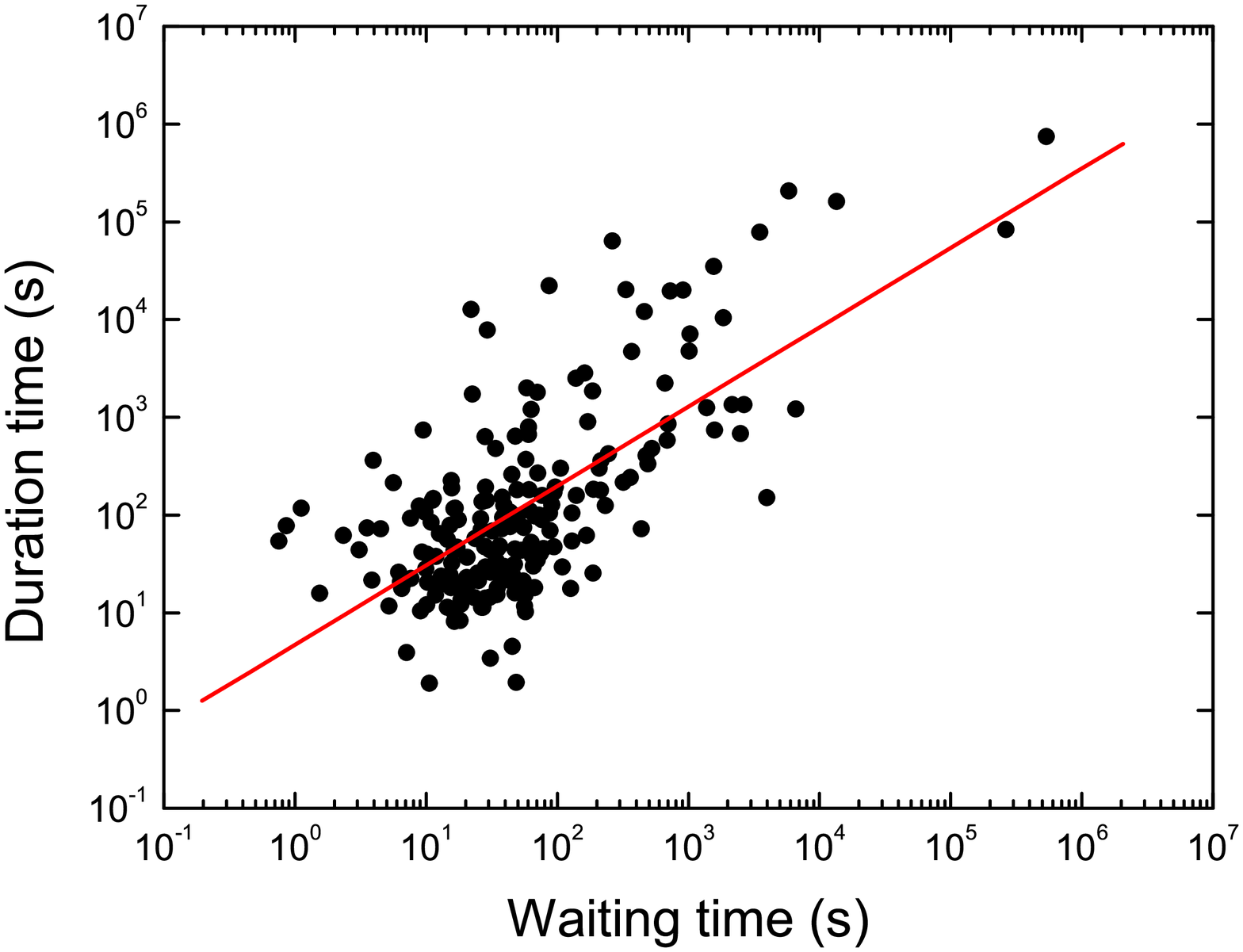}
\includegraphics[angle=0,scale=0.30]{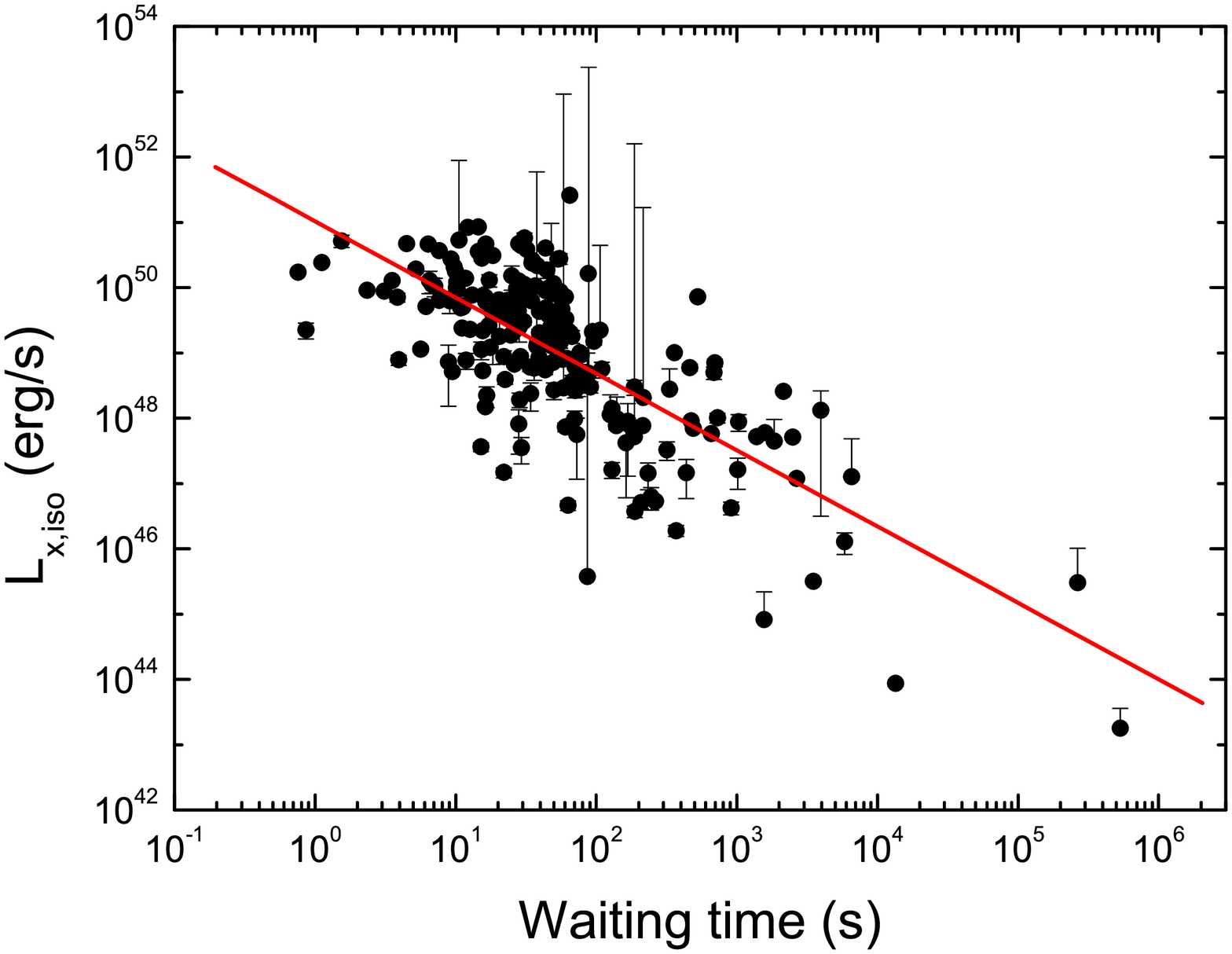}
\includegraphics[angle=0,scale=0.30]{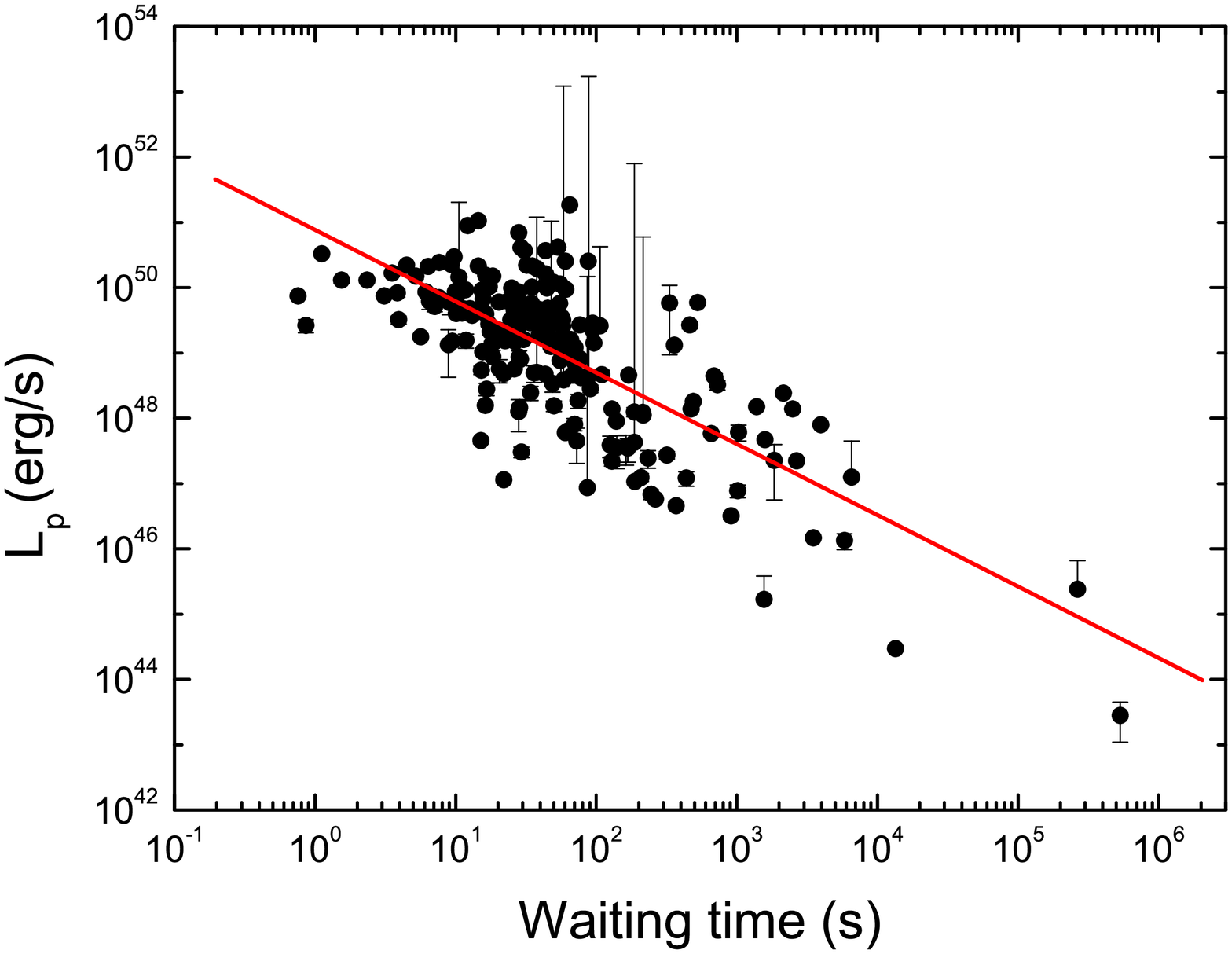}
\caption{The correlations between waiting time and other parameters
of GRB X-ray flares. The symbols have the same meanings as in Figure
4. The best fitting results can be seen in Table 2.}
\end{figure*}

\clearpage
\begin{figure*}
\includegraphics[angle=0,scale=0.30]{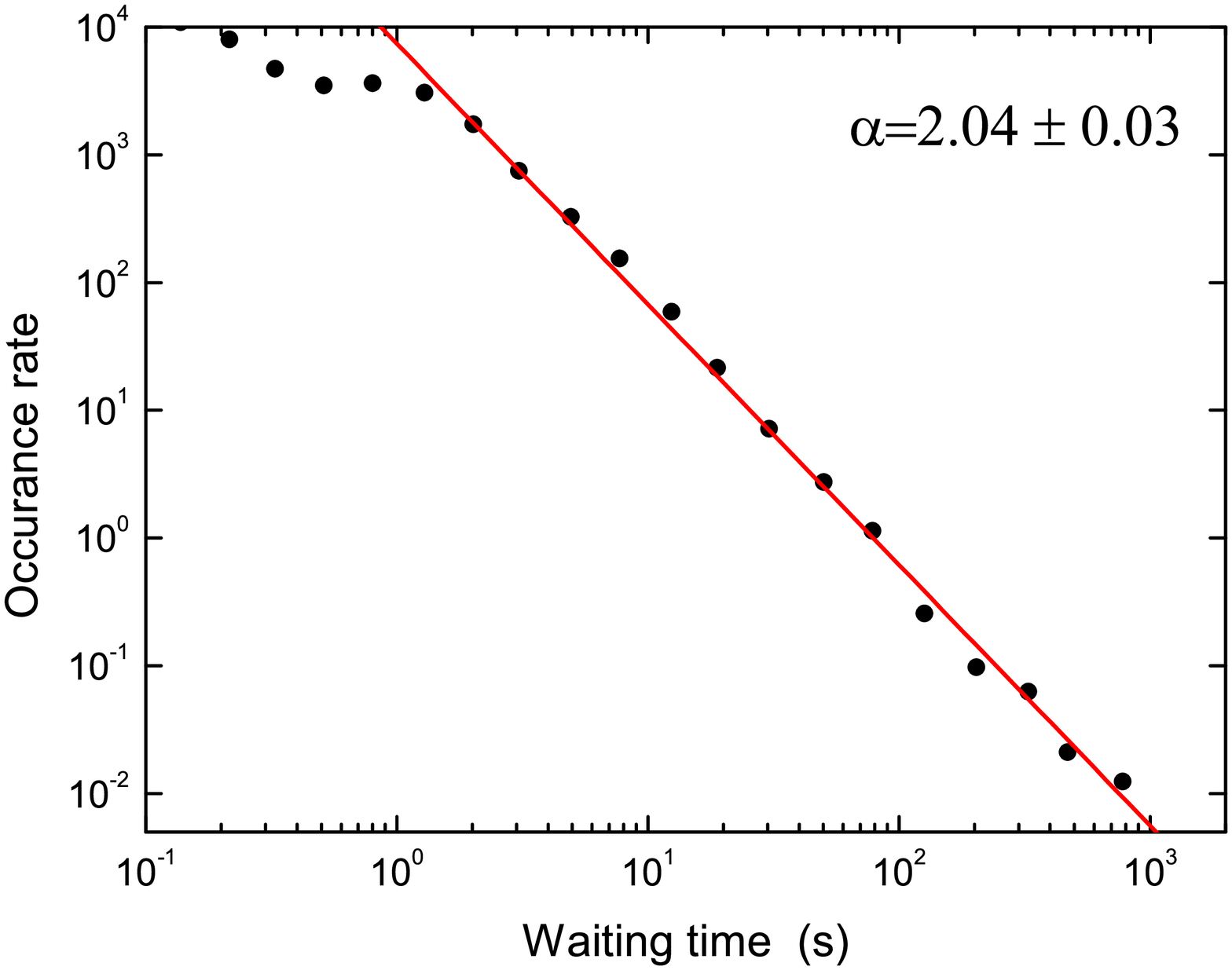}
\includegraphics[angle=0,scale=0.30]{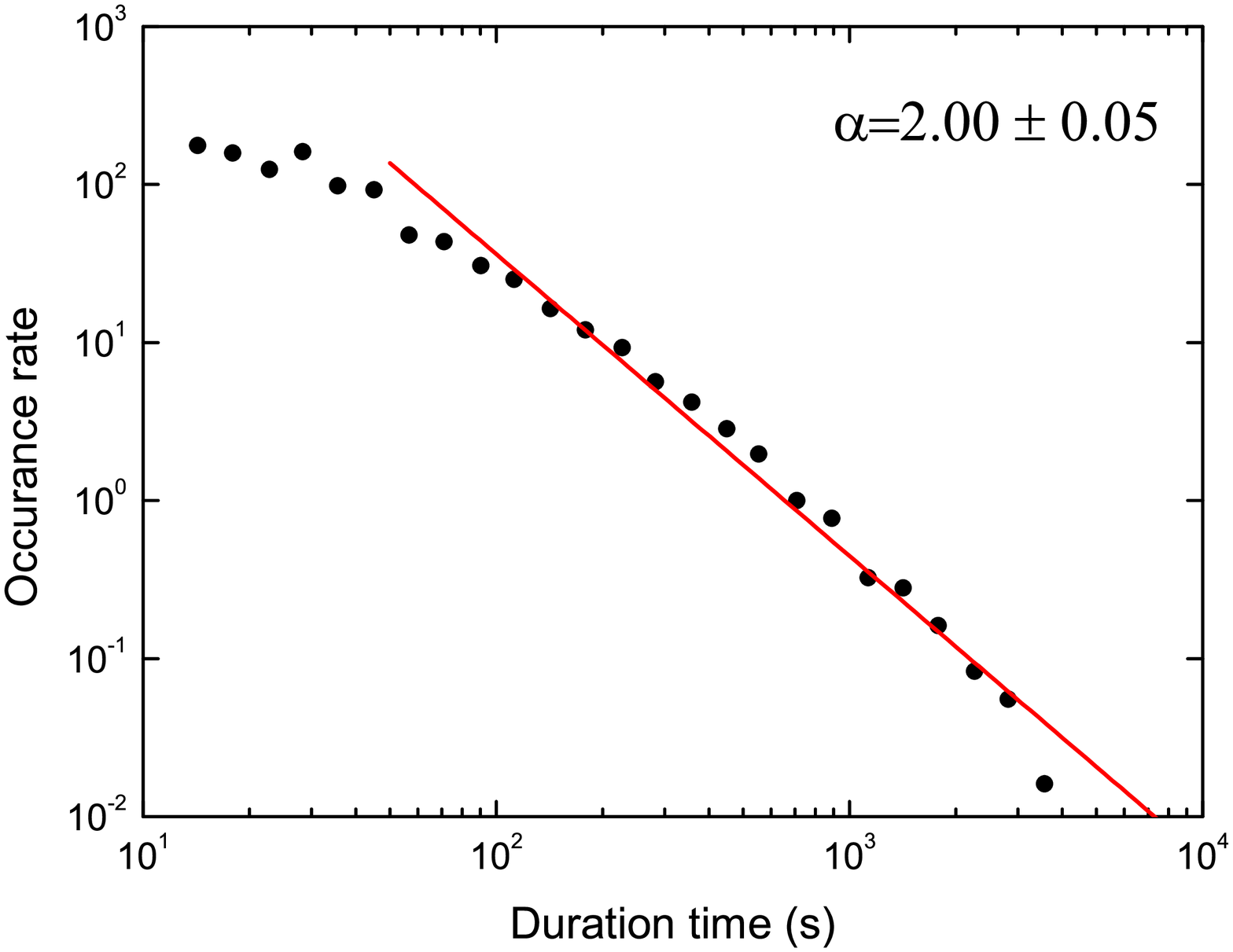}
\includegraphics[angle=0,scale=0.30]{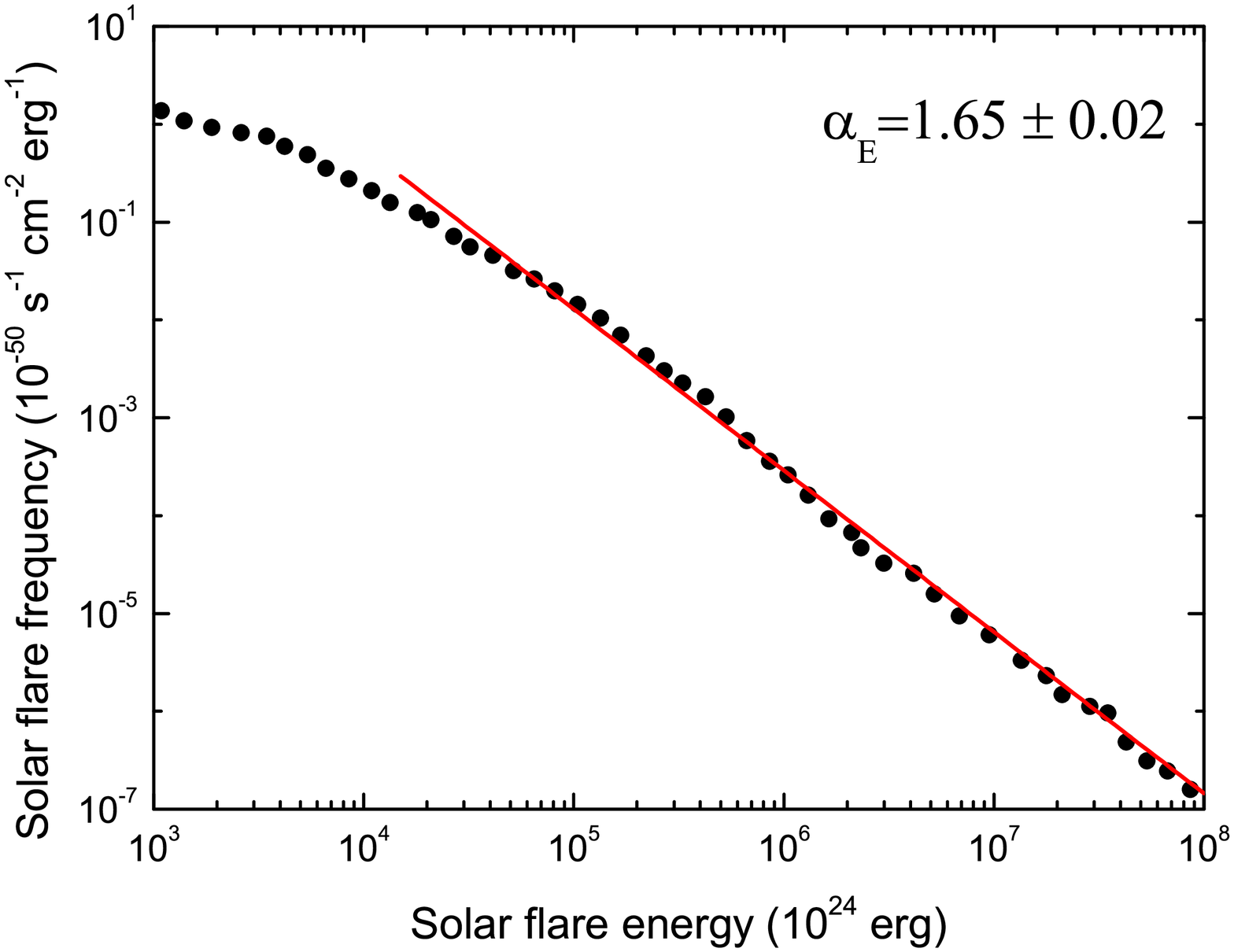}
\includegraphics[angle=0,scale=0.30]{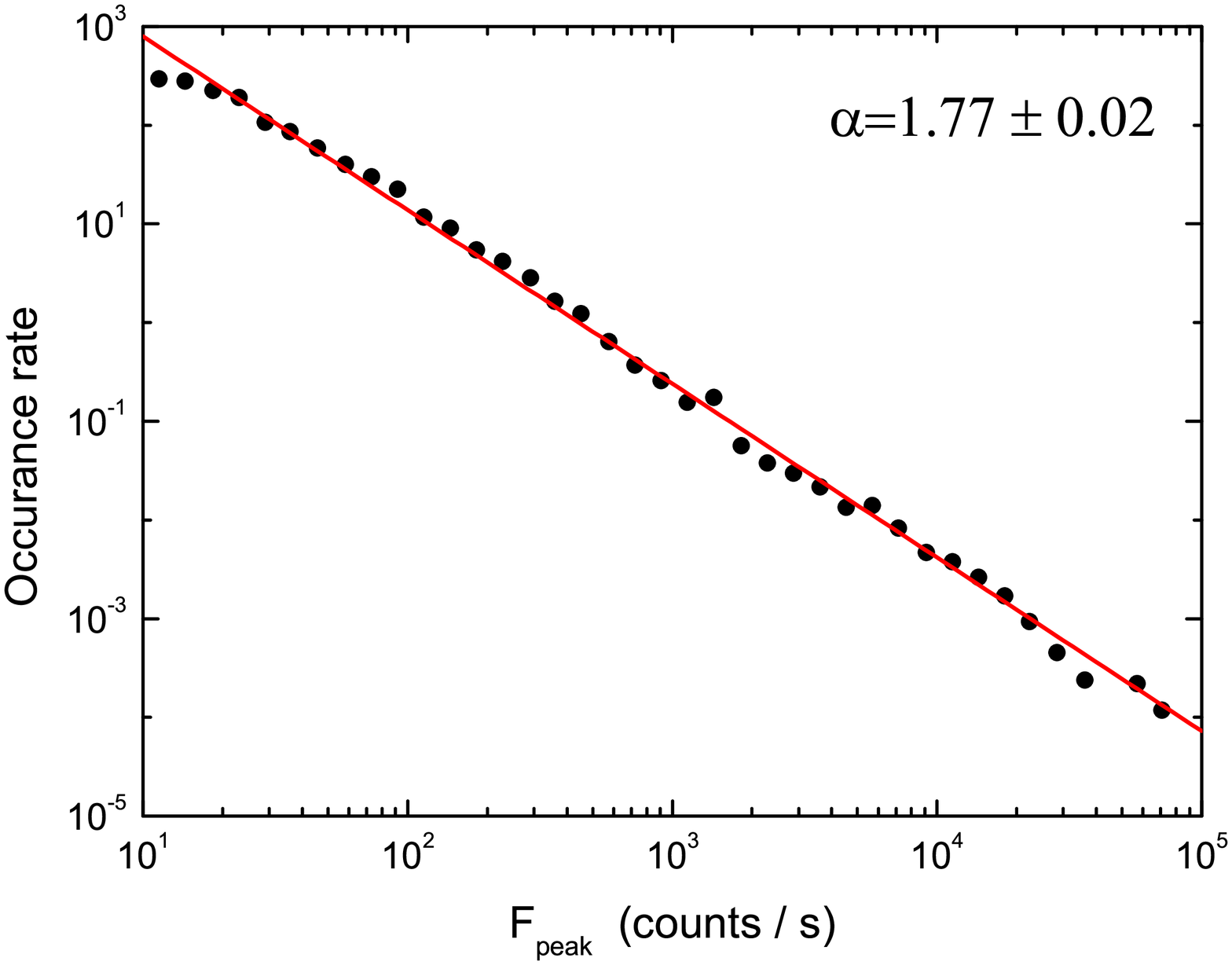}
\caption{The cumulative distributions of solar hard X-ray flares.
11595 solar flares from {\em RHESSI} during 2002-2007 are shown as
black dots (Aschwanden 2011). The best-fit $\alpha$ for the
power-law distributions of waiting time, duration time, energy and
peak flux are $2.04\pm0.03$, $2.00\pm0.05$, $1.65\pm0.02$, and
$1.77\pm0.02$, respectively.}
\end{figure*}

\clearpage
\begin{figure*}
\includegraphics[angle=0,scale=0.40]{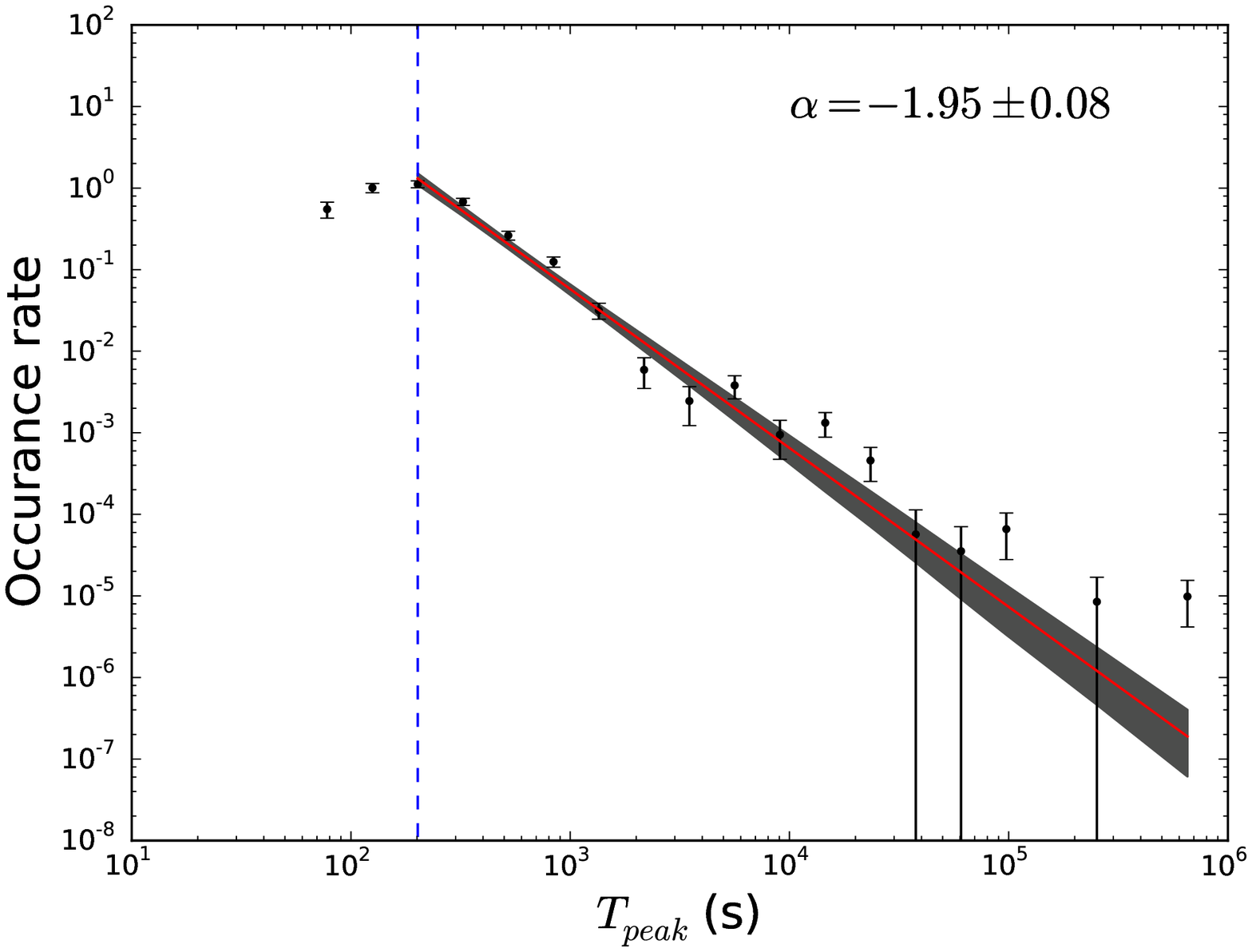}
\includegraphics[angle=0,scale=0.40]{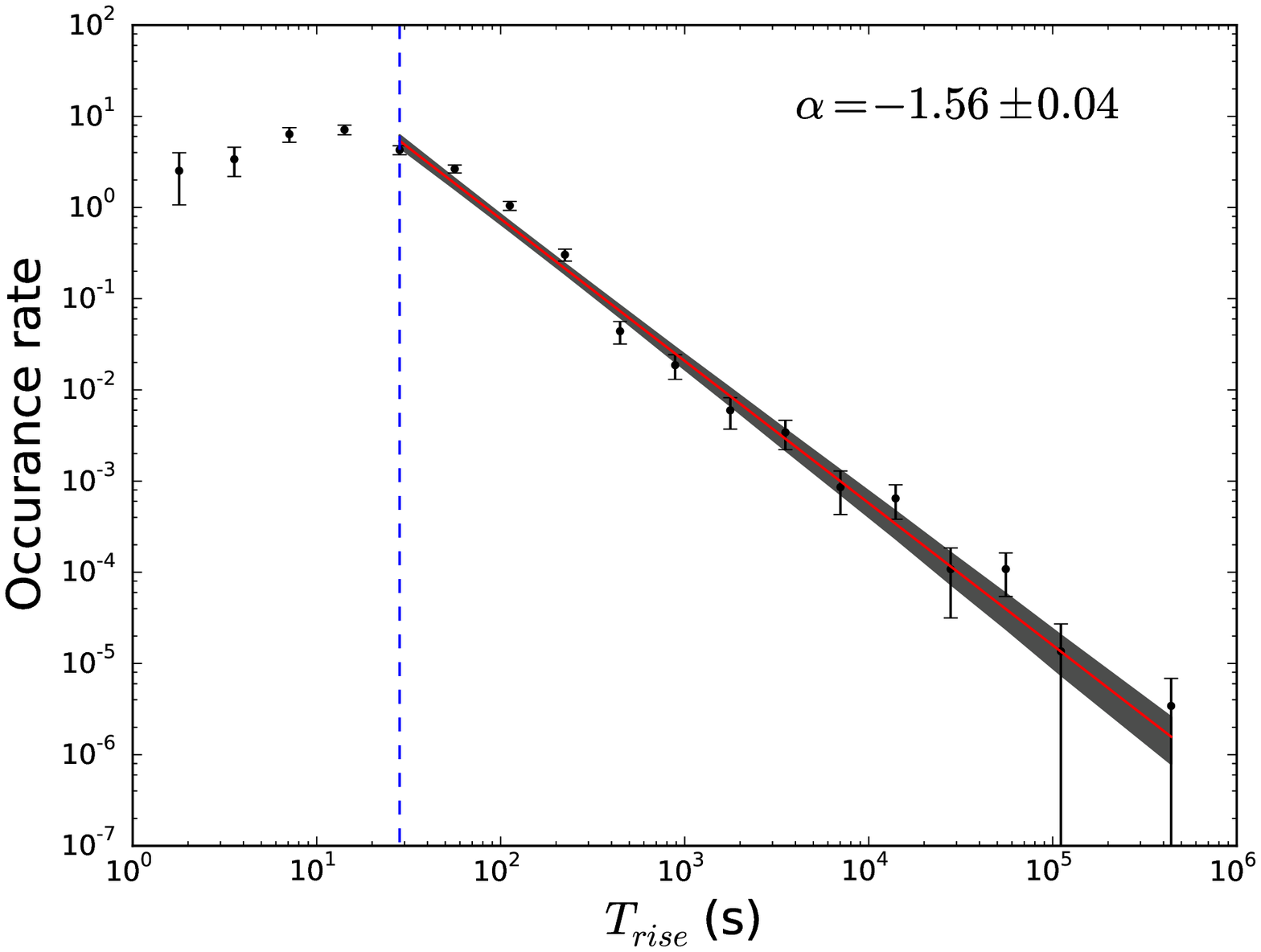}
\includegraphics[angle=0,scale=0.40]{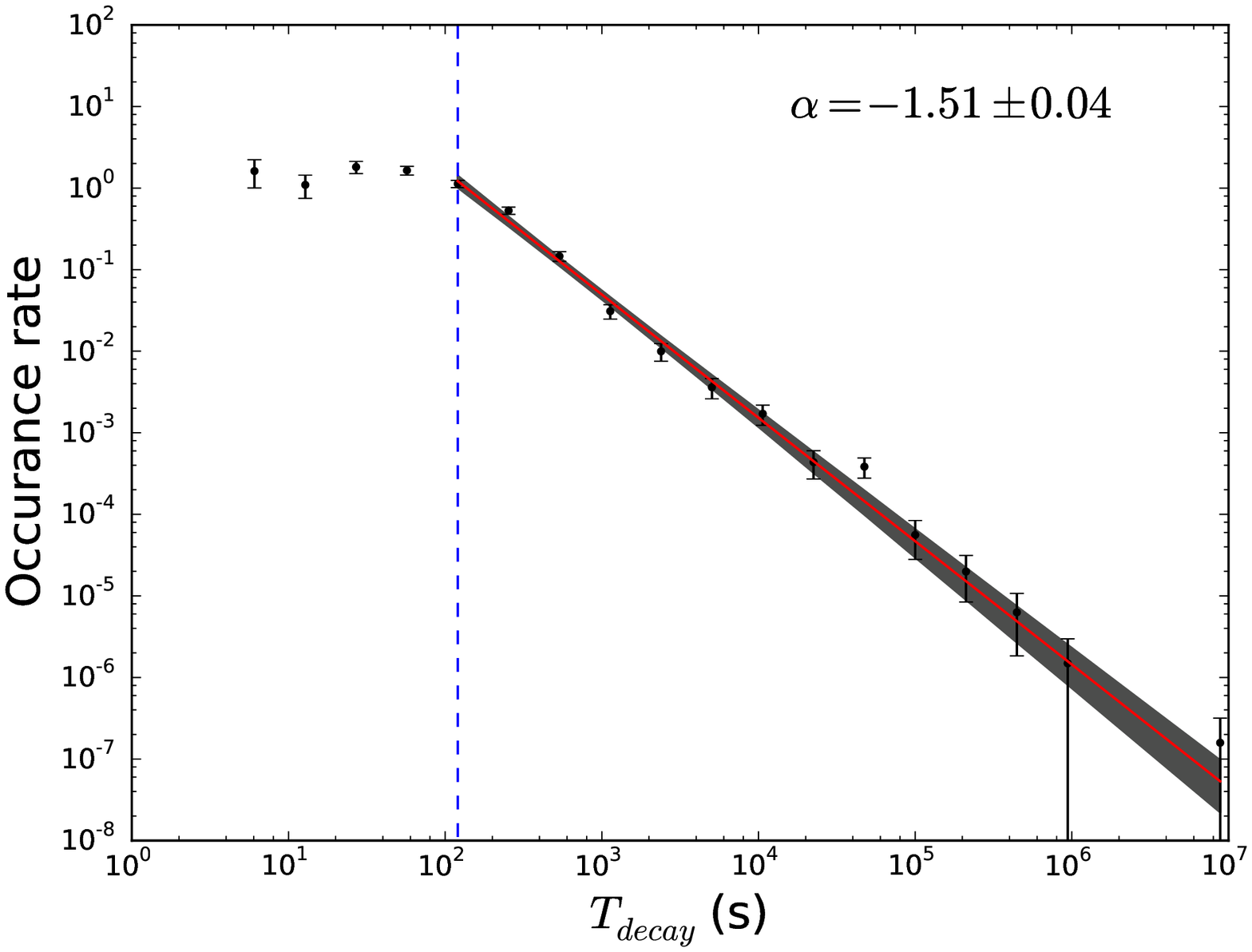}
\includegraphics[angle=0,scale=0.40]{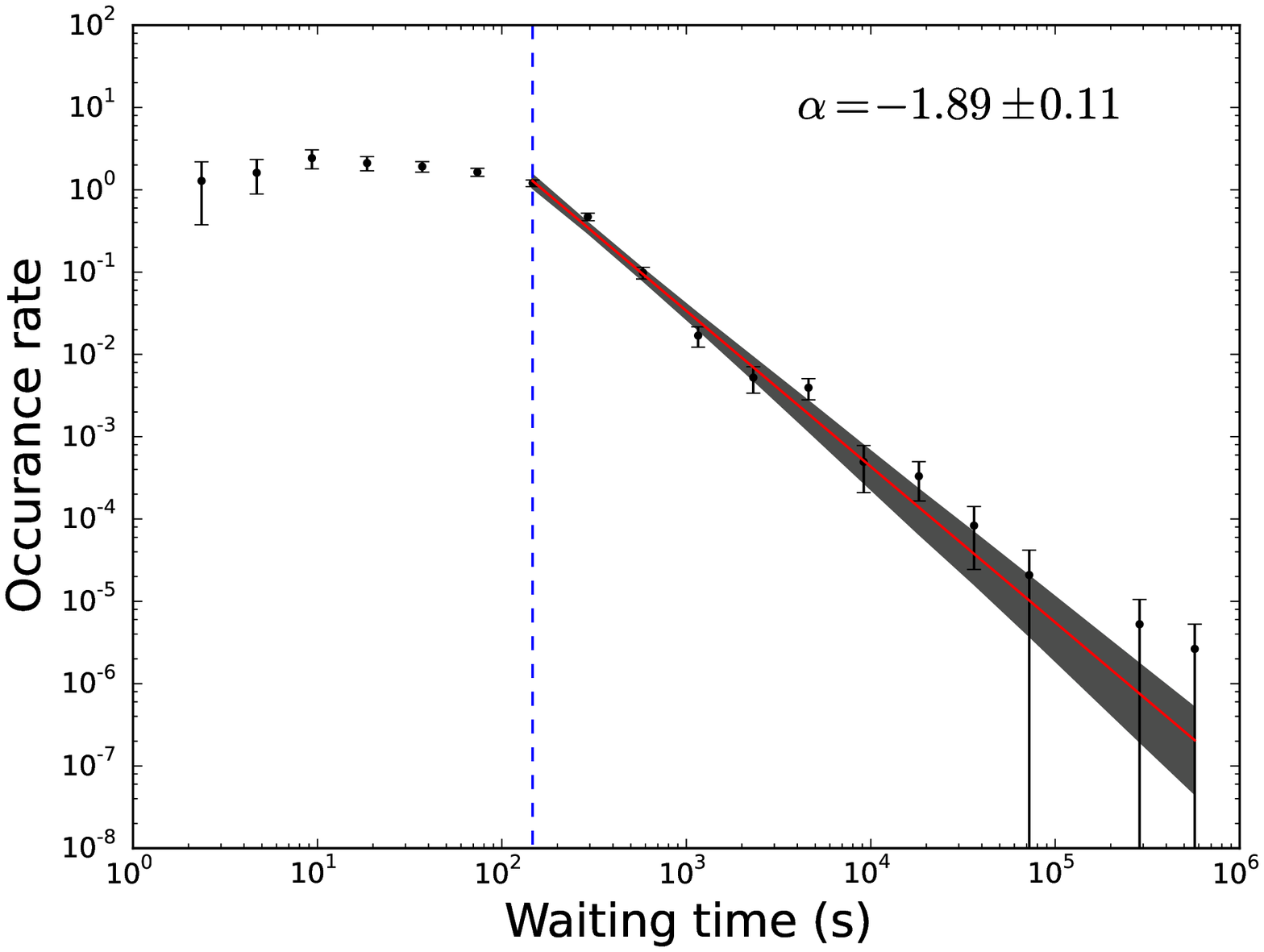}
\includegraphics[angle=0,scale=0.40]{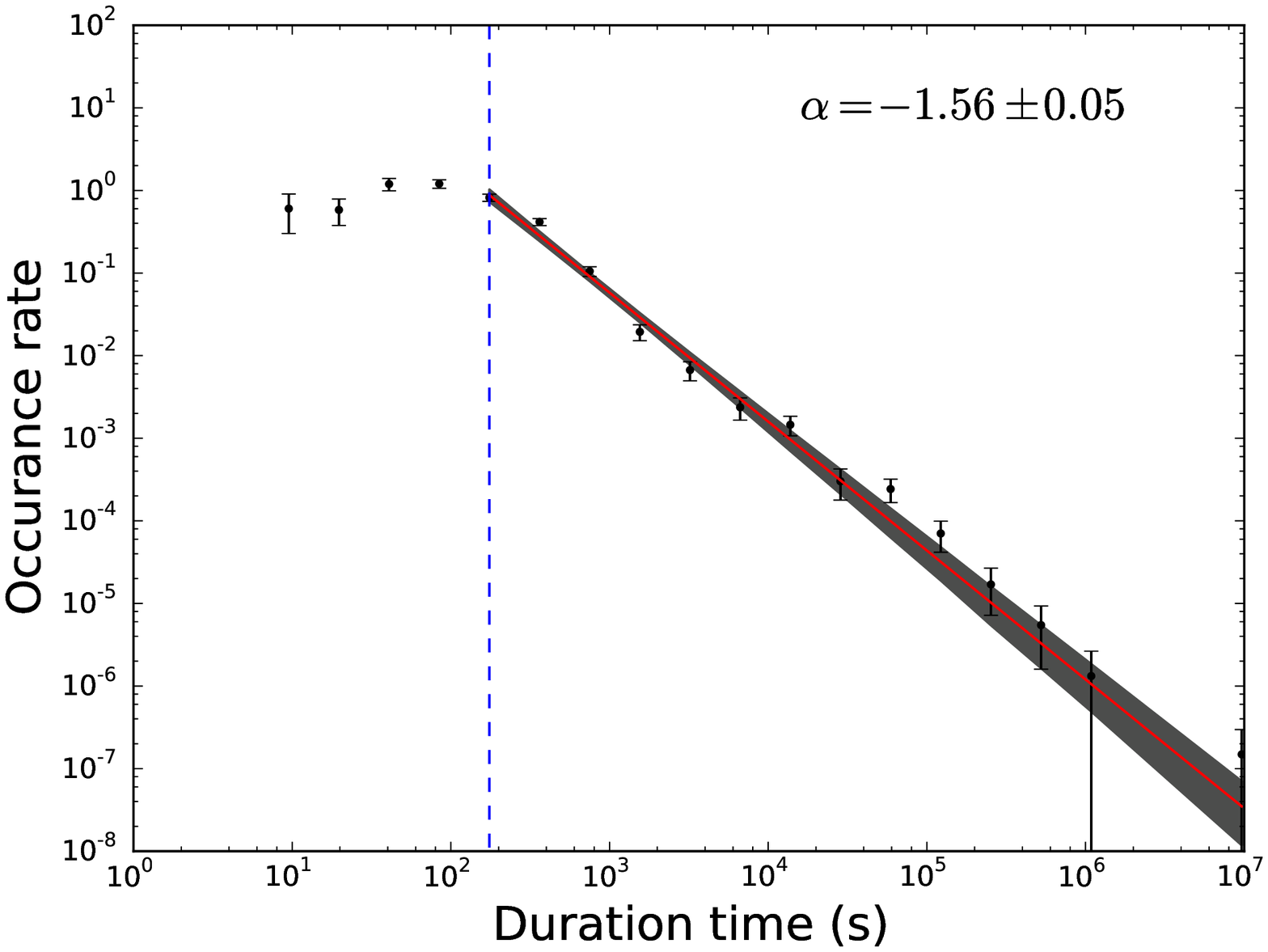}
\includegraphics[angle=0,scale=0.40]{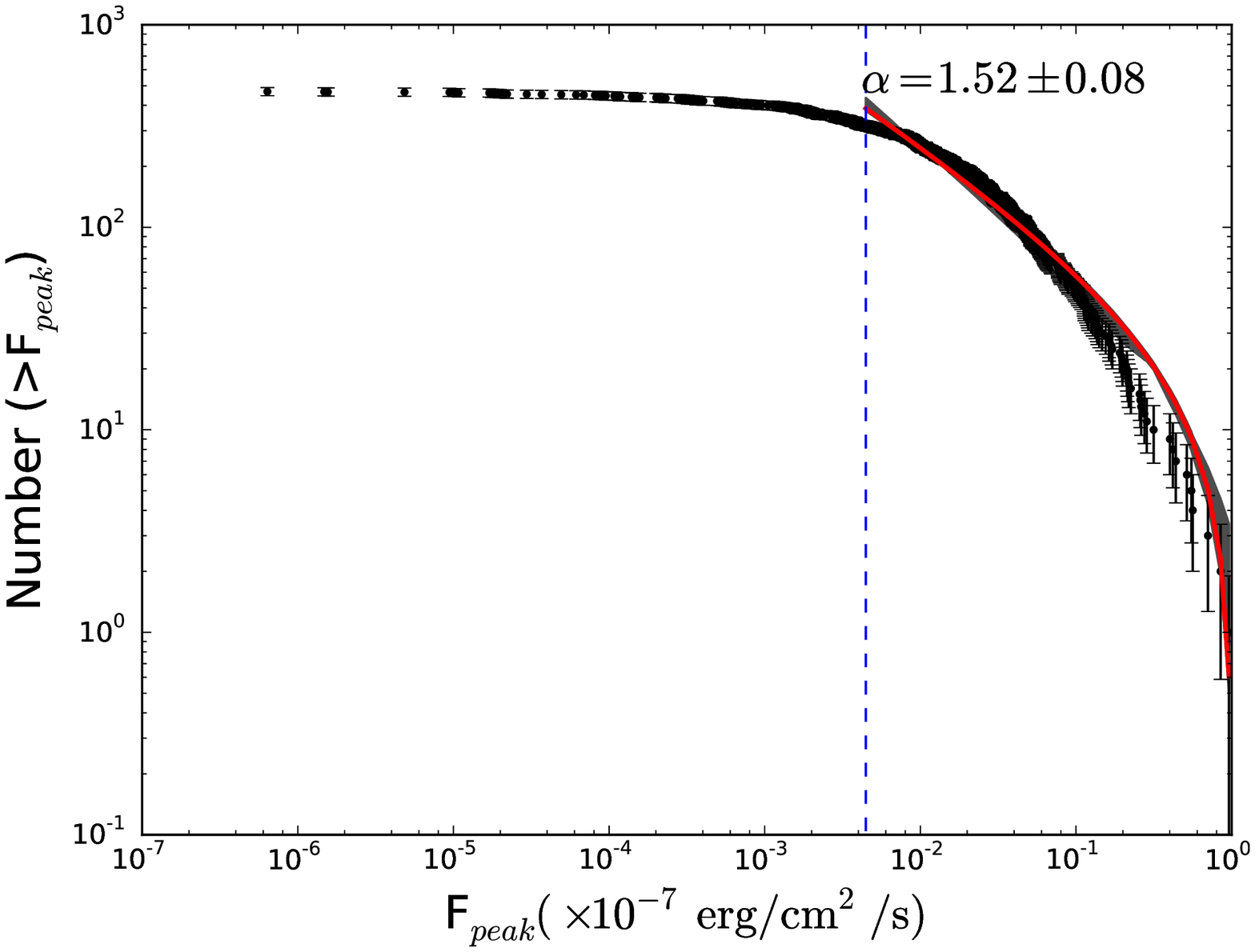}
\caption{The distributions of GRB X-ray flares. The best-fit indices
for the differential distributions of peak time, rise time, decay
time, waiting time and duration time of X-ray flares are
$-1.95\pm0.08$, $-1.56\pm0.04$, $-1.51\pm0.04$, $-1.89\pm0.11$
and$-1.56\pm0.05$ respectively. The gray region shows the 95\%
confidence level. The optimal parameter of the cumulative
distribution for X-ray flare peak flux is $\alpha=1.52\pm0.08$.}
\end{figure*}

\clearpage
\begin{figure*}
\includegraphics[angle=0,scale=0.40]{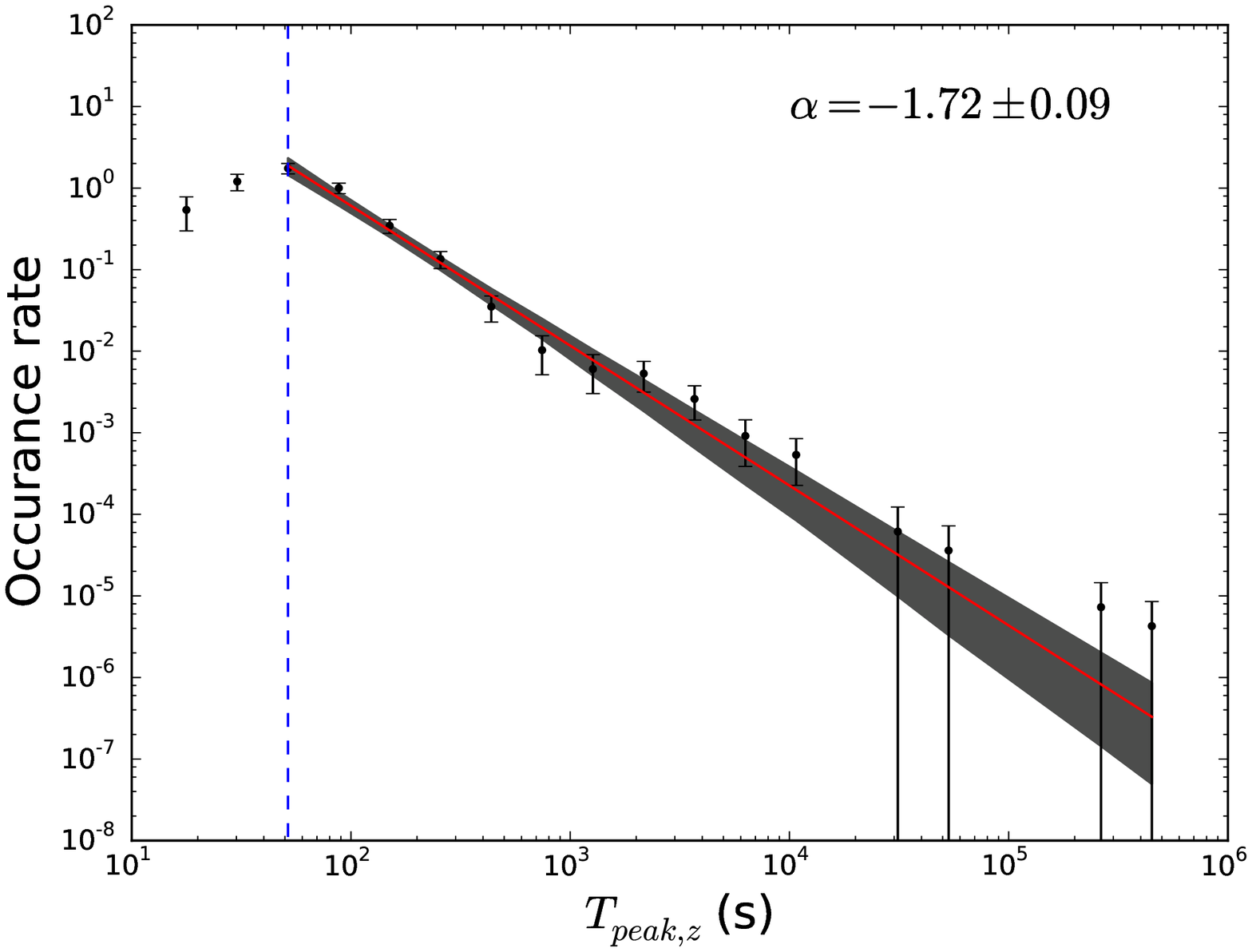}
\includegraphics[angle=0,scale=0.40]{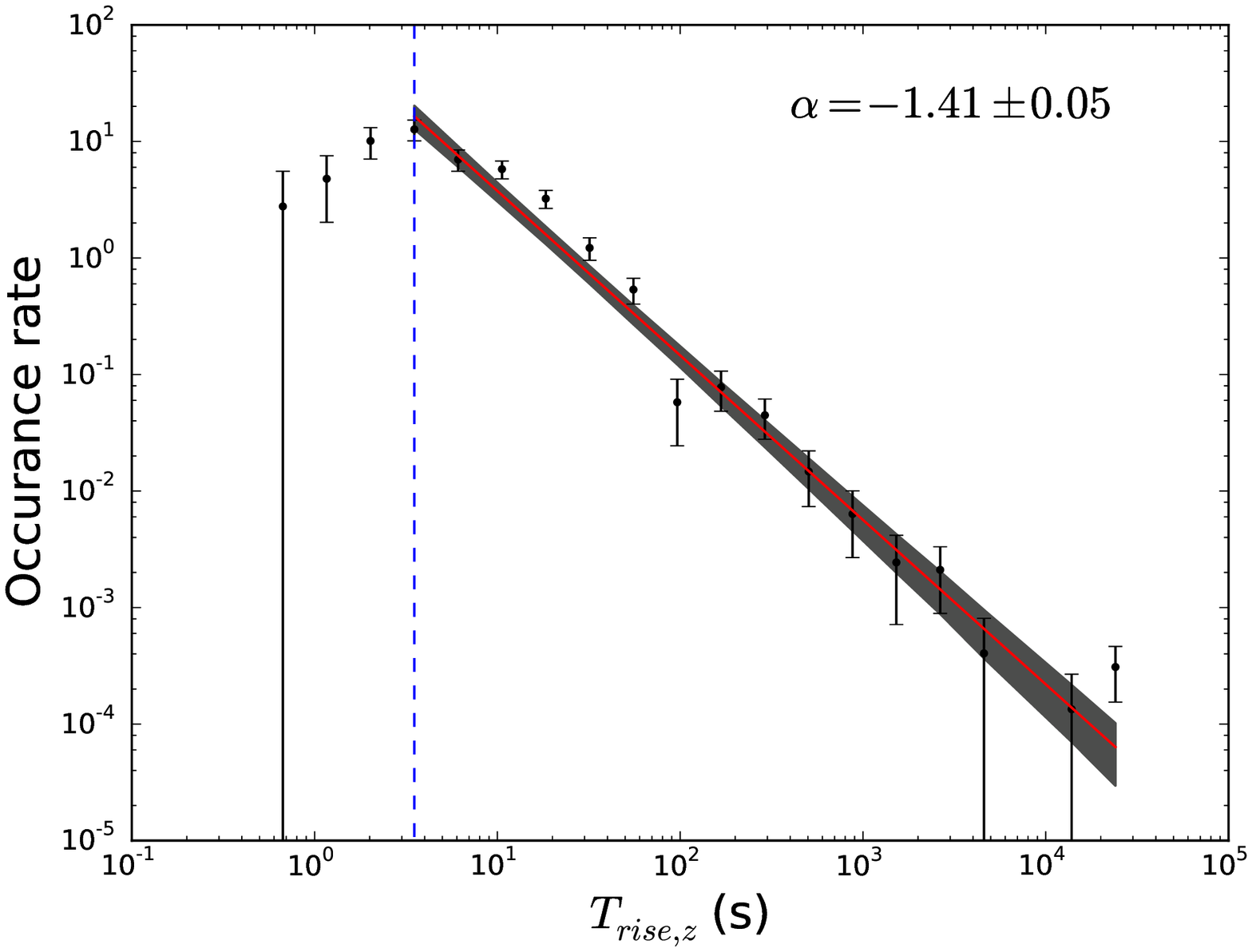}
\includegraphics[angle=0,scale=0.40]{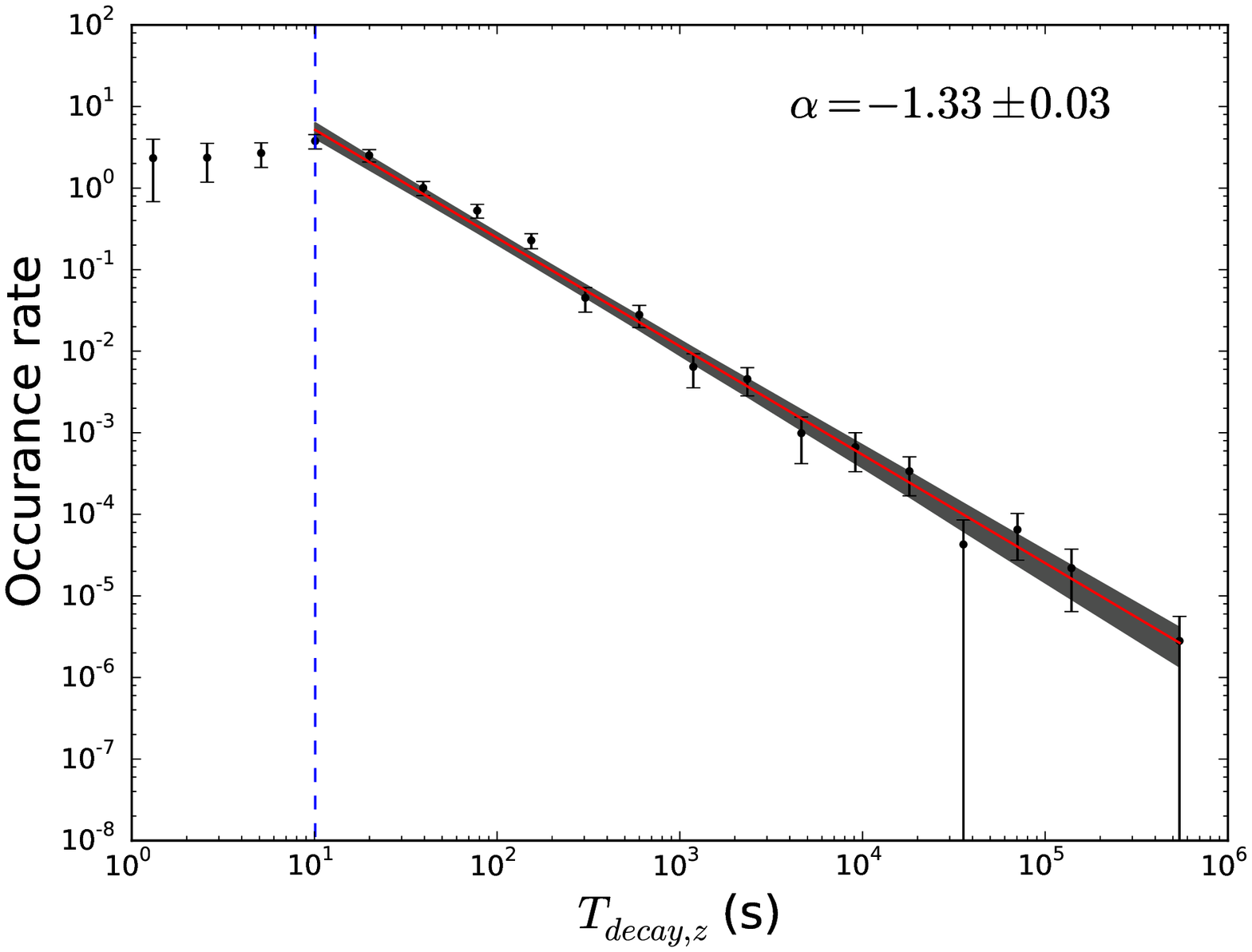}
\includegraphics[angle=0,scale=0.40]{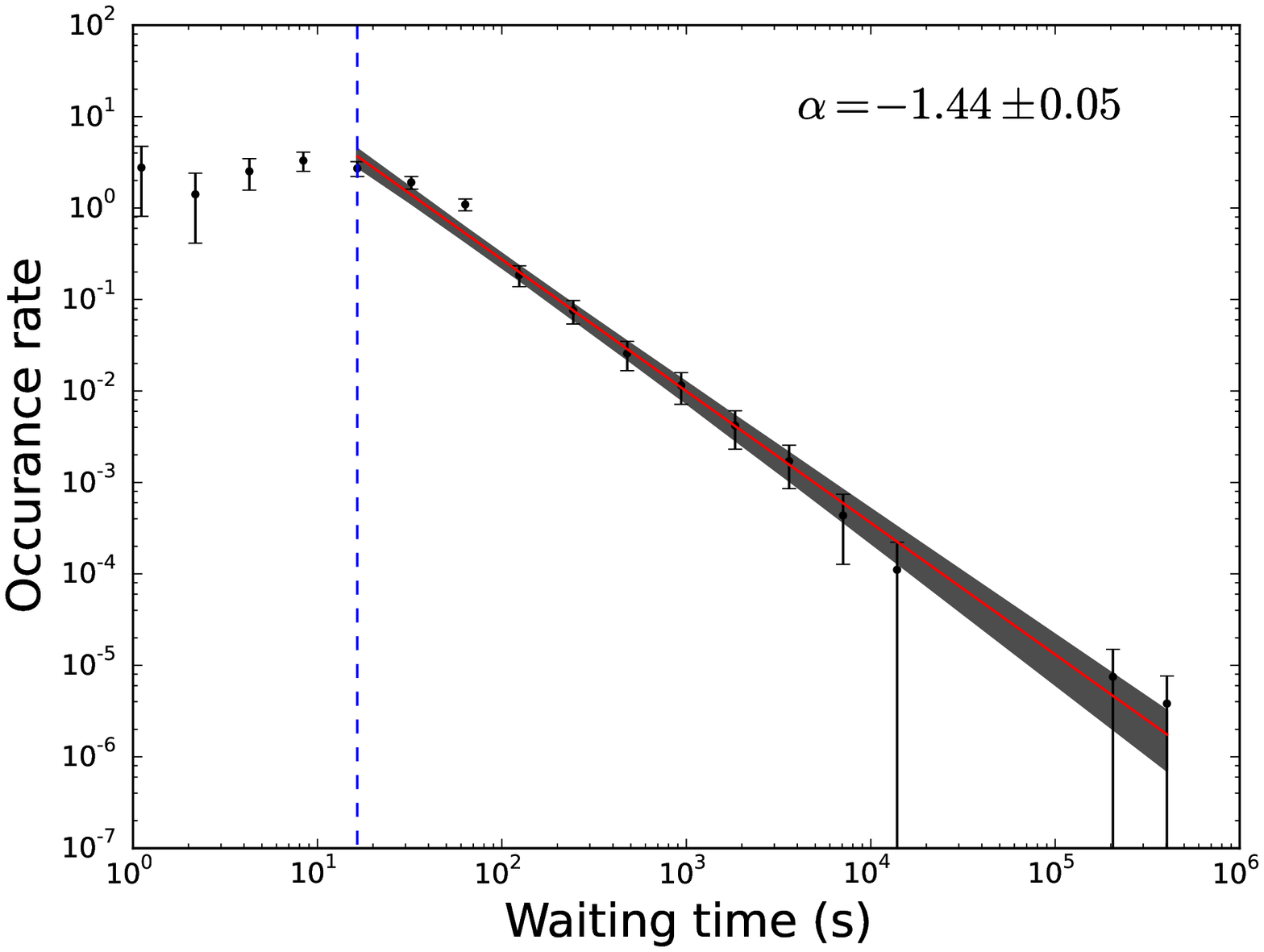}
\includegraphics[angle=0,scale=0.40]{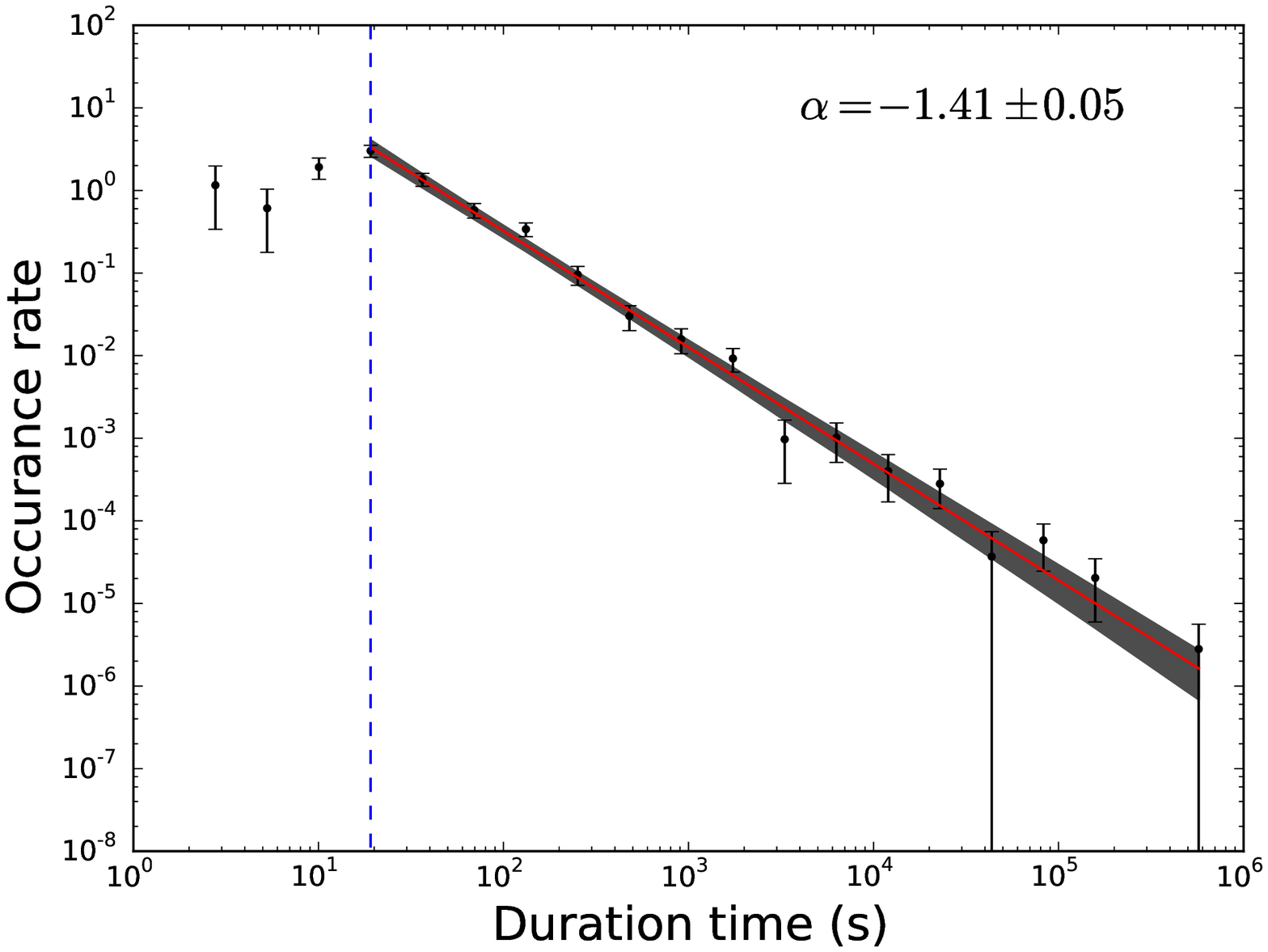}
\includegraphics[angle=0,scale=0.40]{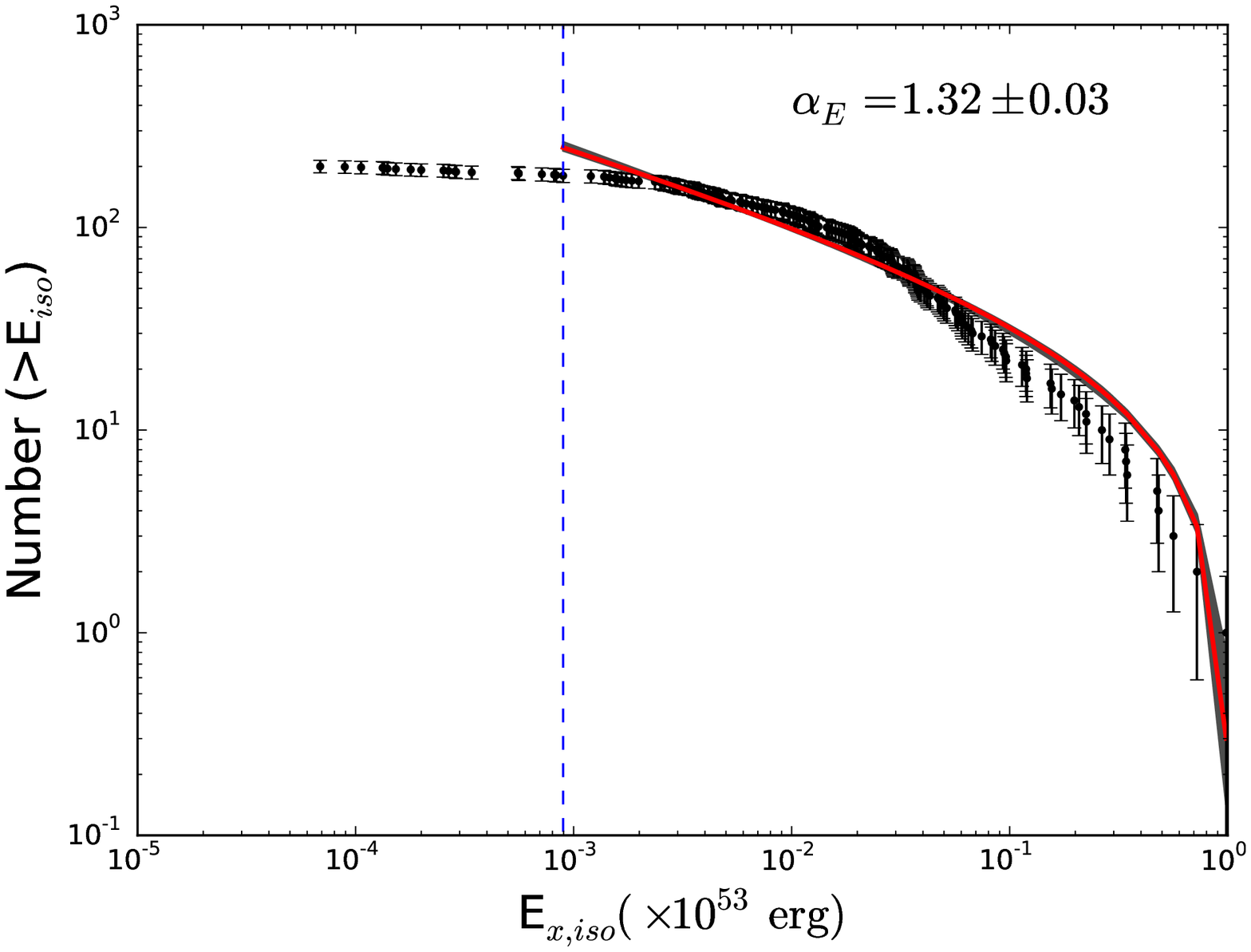}
\caption{The distributions of GRB X-ray flares with redshift. 200
GRB flares are used. The observed times of the flares are
transferred into the source frame. The best-fit indices for the
frequency distributions of peak time, rise time, decay time, waiting
time and duration time of X-ray flares are $-1.72\pm0.09$,
$-1.41\pm0.05$, $-1.33\pm0.03$, $-1.44\pm0.05$ and $-1.41\pm0.05$
respectively. The gray region shows the 95\% confidence level. The
optimal parameter for the cumulative distribution of isotropic
energy of X-ray flares is $\alpha_E=1.32\pm0.03$.}
\end{figure*}

\clearpage
\begin{figure*}
\includegraphics[angle=0,scale=0.30]{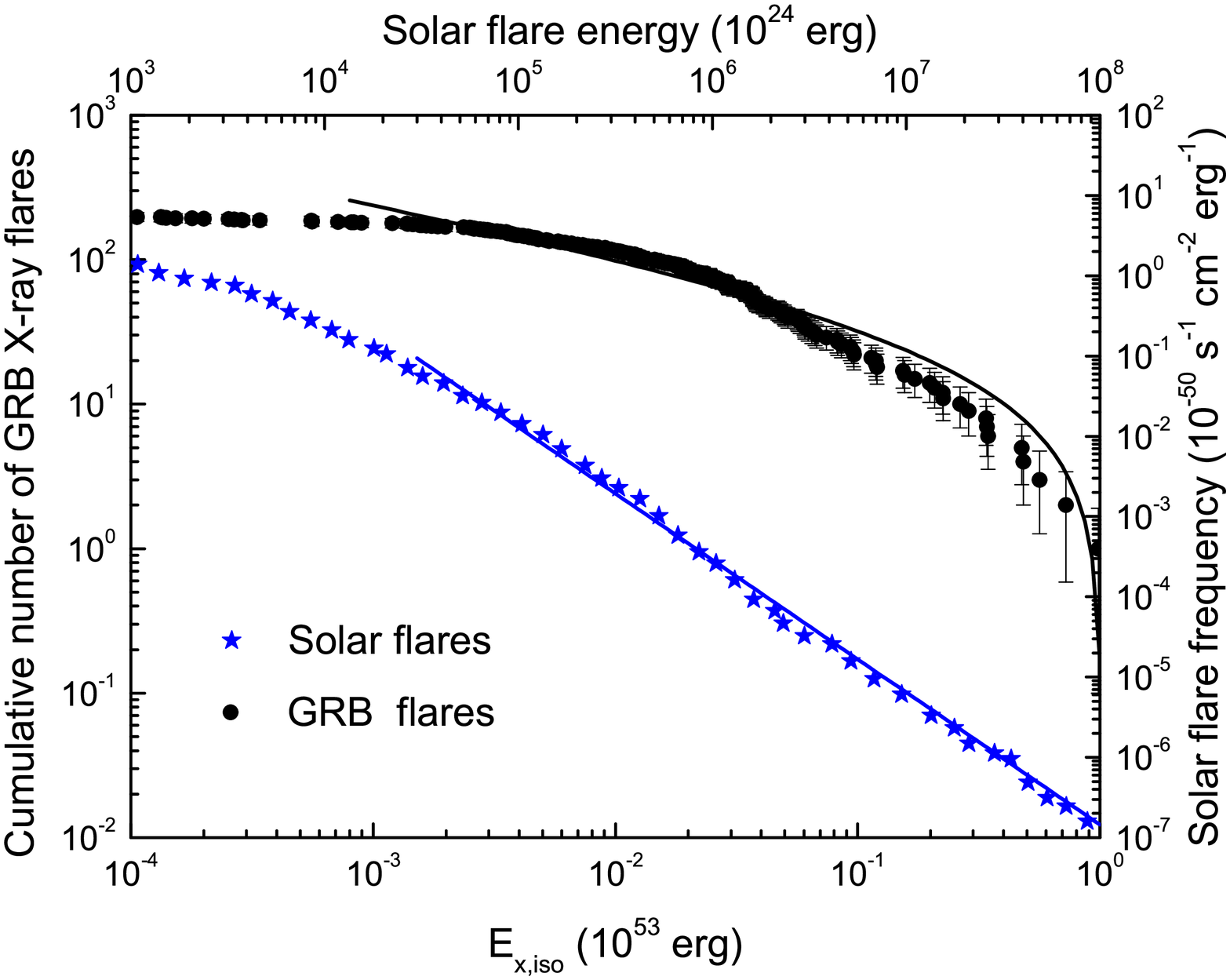}
\includegraphics[angle=0,scale=0.30]{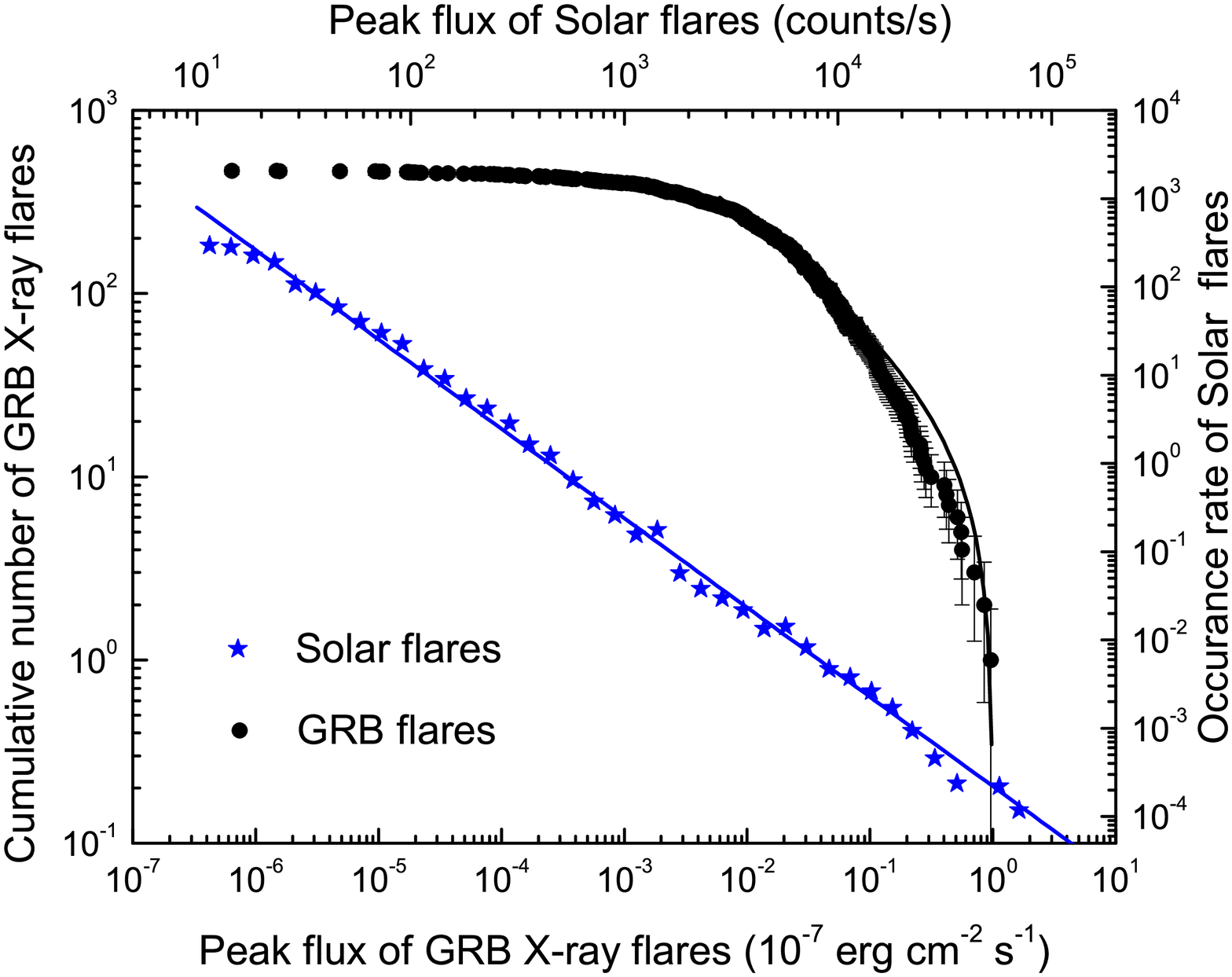}
\includegraphics[angle=0,scale=0.30]{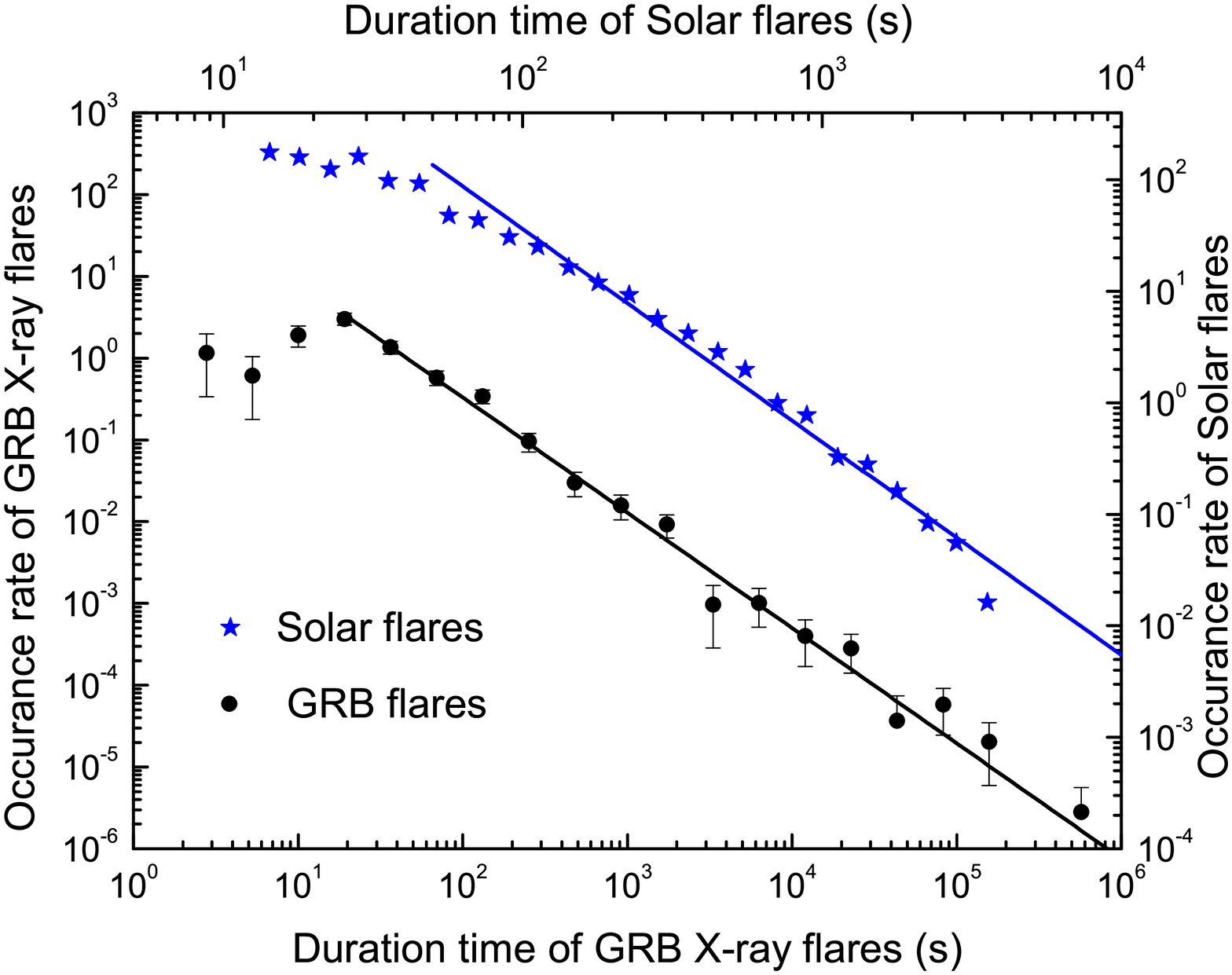}
\includegraphics[angle=0,scale=0.30]{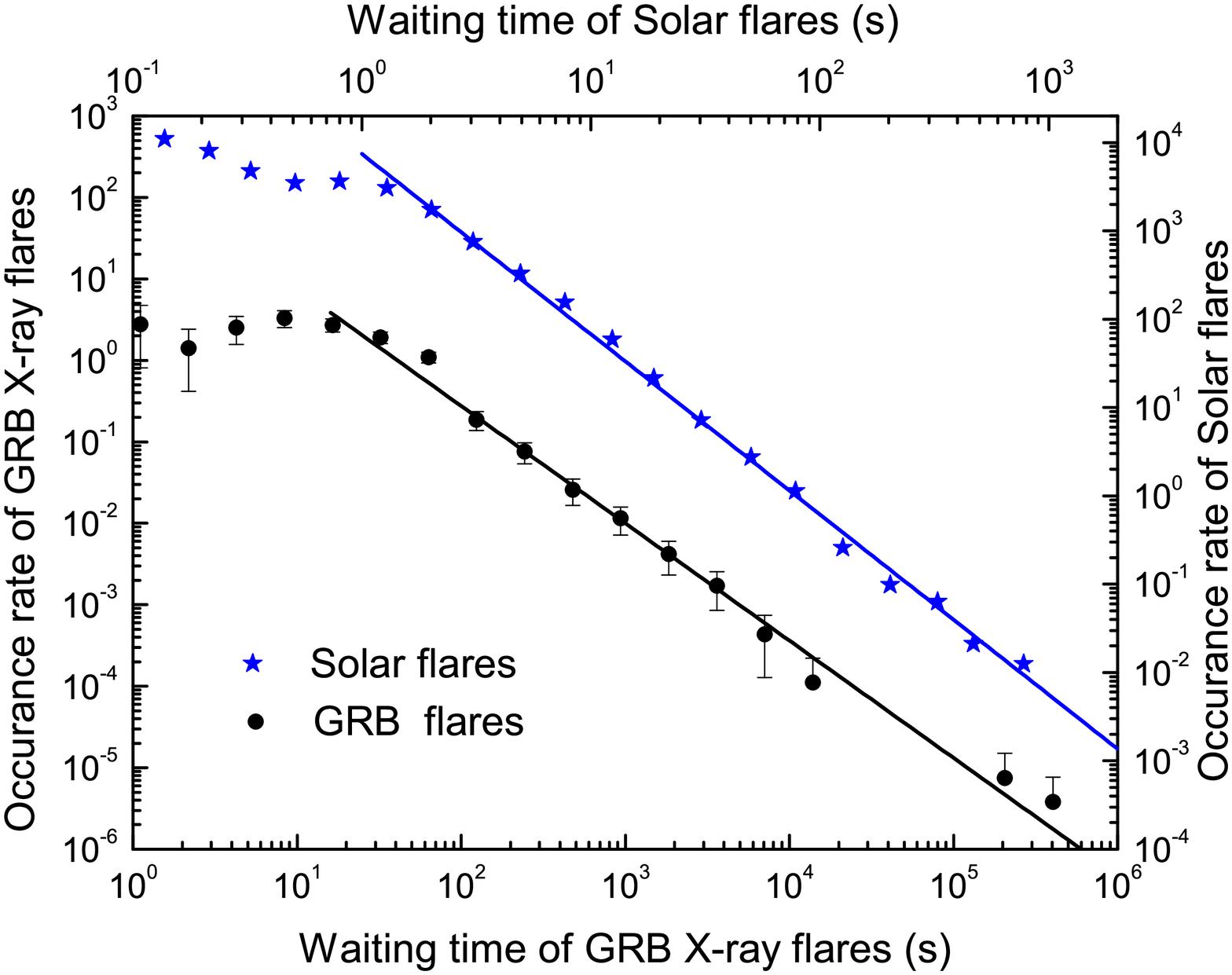}
\caption {The comparison between solar flares (blue stars) and
GRB X-ray flares (black dots). In the up two panels, the
distributions are differential distribution and cumulative
distribution for solar flares and GRB X-ray flares, respectively.
The best-fitting indices can be found in Figures 7, 8 and 9.}
\end{figure*}

%&&&&&&&&&&&&&&&&&&&&&&&&&&&&&&&&&&&&&&&&&&&&&&&&&&&&&&&&&&&&&&&&&&&&&&&&&&&&&&&&&&&&&&&&&&&&&&&&&&&&&&&&&&&&&&&

%\clearpage
%


\begin{thebibliography}{99}

\bibitem[Abramowski et al.(2012)]{2012ApJ...746..151A} Abramowski, A.,
Acero, F., Aharonian, F., et al.\ 2012, \apj, 746, 151


\bibitem[Aschwanden 2011]{asc11} Aschwanden, M. J., 2011, Self-Organized Criticality in Astrophysics:
The Statistics of Nonlinear Processes in the Universe, Springer-Verlag: Berlin


\bibitem[Aschwanden(2012)]{2012A&A...539A...2A} Aschwanden, M.~J.\ 2012, \aap, 539, A2


\bibitem[Baganoff et al.(2001)]{2001Natur.413...45B} Baganoff, F.~K.,
Bautz, M.~W., Brandt, W.~N., et al.\ 2001, \nat, 413, 45

\bibitem[Bak et al.(1988)]{1988PhRvA..38..364B} Bak, P., Tang, C.,
\& Wiesenfeld, K.\ 1988, \pra, 38, 364


\bibitem[Bak et al.(1987)]{1987PhRvL..59..381B} Bak, P., Tang, C.,
\& Wiesenfeld, K.\ 1987, Physical Review Letters, 59, 381



\bibitem[Bernardini et
al.(2011)]{2011A&A...526A..27B} Bernardini, M.~G., Margutti, R., Chincarini, G., Guidorzi, C., \& Mao, J.\ 2011, \aap, 526, A27


\bibitem[Burrows et al.(2011)]{2011Natur.476..421B} Burrows, D.~N., Kennea,
J.~A., Ghisellini, G., et al.\ 2011, \nat, 476, 421


\bibitem[Burrows et al.(2005)]{2005Sci...309.1833B} Burrows, D.~N., Romano,
P., Falcone, A., et al.\ 2005, Science, 309, 1833


\bibitem[Campana et
al.(2006)]{2006A&A...454..113C} Campana, S., Tagliaferri, G., Lazzati, D., et al.\ 2006, \aap, 454, 113

\bibitem[Chapman et al.(1998)]{1998GeoRL..25.2397C} Chapman, S.~C.,
Watkins, N.~W., Dendy, R.~O., Helander, P.,
\& Rowlands, G.\ 1998, \grl, 25, 2397


\bibitem[Charbonneau
\& MacGregor(2001)]{2001ApJ...559.1094C} Charbonneau, P., \& MacGregor, K.~B.\ 2001, \apj, 559, 1094


\bibitem[Chincarini et al.(2010)]{2010MNRAS.406.2113C} Chincarini, G., Mao,
J., Margutti, R., et al.\ 2010, \mnras, 406, 2113


\bibitem[Chincarini et al.(2007)]{2007ApJ...671.1903C} Chincarini, G.,
Moretti, A., Romano, P., et al.\ 2007, \apj, 671, 1903


\bibitem[Cliver et al.(2012)]{2012ApJ...756L..29C} Cliver, E.~W., Ling,
A.~G., Belov, A., \& Yashiro, S.\ 2012, \apjl, 756, L29

\bibitem[Crosby et al.(1996)]{1996SoPh..167..333C} Crosby, N., Vilmer, N.,
Lund, N., Klein, K.-L., \& Sunyaev, R.\ 1996, \solphys, 167, 333



\bibitem[Dai et al.(2006)]{2006Sci...311.1127D} Dai, Z.~G., Wang, X.~Y.,
Wu, X.~F., \& Zhang, B.\ 2006, Science, 311, 1127


\bibitem[De Young(1991)]{1991Sci...252..389D} De Young, D.~S.\ 1991,
Science, 252, 389


\bibitem[Evans et
al.(2007)]{2007A&A...469..379E} Evans, P.~A., Beardmore, A.~P., Page, K.~L., et al.\ 2007, \aap, 469, 379

\bibitem[Evans et al.(2009)]{2009MNRAS.397.1177E} Evans, P.~A., Beardmore,
A.~P., Page, K.~L., et al.\ 2009, \mnras, 397, 1177



\bibitem[Falcone et al.(2006)]{2006ApJ...641.1010F} Falcone, A.~D.,
Burrows, D.~N., Lazzati, D., et al.\ 2006, \apj, 641, 1010


\bibitem[Falcone et al.(2007)]{2007ApJ...671.1921F} Falcone, A.~D., Morris,
D., Racusin, J., et al.\ 2007, \apj, 671, 1921


\bibitem[Fan
\& Wei(2005)]{2005MNRAS.364L..42F} Fan, Y.~Z., \& Wei, D.~M.\ 2005, \mnras, 364, L42


\bibitem[Giannios(2006)]{2006A&A...455L...5G} Giannios, D.\ 2006, \aap, 455, L5


\bibitem[Guidorzi et al.(2015)]{2015ApJ...801...57G} Guidorzi, C.,
Dichiara, S., Frontera, F., et al.\ 2015, \apj, 801, 57


\bibitem[Harko et
al.(2015)]{2015Ap&SS.357...84H} Harko, T., Mocanu, G., \& Stroia, N.\ 2015, \apss, 357, 84


\bibitem[Harris et al.(2009)]{2009ApJ...699..305H} Harris, D.~E., Cheung,
C.~C., Stawarz, {\L}., Biretta, J.~A.,
\& Perlman, E.~S.\ 2009, \apj, 699, 305


\bibitem[Hou et al.(2014)]{2014ApJ...785..113H} Hou, S.~J., Geng, J.~J.,
Wang, K., et al.\ 2014, \apj, 785, 113


\bibitem[Jia et al.(2015)]{2015arXiv150904871J} Jia, L.-W., Uhm, Z.~L.,
\& Zhang, B.\ 2015, arXiv:1509.04871


\bibitem[King et al.(2005)]{2005ApJ...630L.113K} King, A., O'Brien, P.~T.,
Goad, M.~R., et al.\ 2005, \apjl, 630, L113


\bibitem[Klu{\'z}niak
\& Ruderman(1998)]{1998ApJ...505L.113K} Klu{\'z}niak, W., \& Ruderman, M.\ 1998, \apjl, 505, L113



\bibitem[Kumar(1999)]{1999ApJ...523L.113K} Kumar, P.\ 1999, \apjl, 523,
L113

\bibitem[Kumar
\& Panaitescu(2000)]{2000ApJ...541L...9K} Kumar, P., \& Panaitescu,
A.\ 2000, \apjl, 541, L51



\bibitem[Kobayashi et al.(1997)]{1997ApJ...490...92K} Kobayashi, S., Piran,
T., \& Sari, R.\ 1997, \apj, 490, 92


\bibitem[Lei et al.(2013)]{2013ApJ...765..125L} Lei, W.-H., Zhang, B.,
\& Liang, E.-W.\ 2013, \apj, 765, 125


\bibitem[Li et al.(2012)]{2012ApJ...758...27L} Li, L., Liang, E.-W., Tang,
Q.-W., et al.\ 2012, \apj, 758, 27

\bibitem[Liang et al.(2006)]{2006ApJ...646..351L} Liang, E.-W., Zhang, B.,
O'Brien, P. T., et al.\ 2006, \apj, 646, 351

\bibitem[Liang et al.(2010)]{2010ApJ...725.2209L} Liang, E.-W., Yi, S.-X.,
Zhang, J., et al.\ 2010, \apj, 725, 2209


\bibitem[Lu
\& Hamilton(1991)]{1991ApJ...380L..89L} Lu, E.~T., \& Hamilton, R.~J.\ 1991, \apjl, 380, L89

\bibitem[Margutti et al.(2011)]{2011MNRAS...417} Margutti, M., et al.\ 2011, \mnras, 417,
2144

\bibitem[Meier et al.(2001)]{2001Sci...291...84M} Meier, D.~L., Koide, S.,
\& Uchida, Y.\ 2001, Science, 291, 84


\bibitem[Mirabel
\& Rodr{\'{\i}}guez(1999)]{1999ARA&A..37..409M} Mirabel, I.~F., \& Rodr{\'{\i}}guez, L.~F.\ 1999, \araa, 37, 409


\bibitem[Morales
\& Charbonneau(2008)]{2008ApJ...682..654M} Morales, L., \& Charbonneau, P.\ 2008, \apj, 682, 654


\bibitem[Negoro et al.(1995)]{1995ApJ...452L..49N} Negoro, H., Kitamoto,
S., Takeuchi, M., \& Mineshige, S.\ 1995, \apjl, 452, L49


\bibitem[Neilsen et al.(2013)]{2013ApJ...774...42N} Neilsen, J., Nowak,
M.~A., Gammie, C., et al.\ 2013, \apj, 774, 42


\bibitem[Nousek et al.(2006)]{2006ApJ...642..389N} Nousek, J.~A.,
Kouveliotou, C., Grupe, D., et al.\ 2006, \apj, 642, 389

\bibitem[Omori(1895)]{asc11} Omori, F.\ 1895, J. Coll. Sci. Imper. Univ. Tokyo, 7, 111



\bibitem[Parker(1957)]{1957PhRv..107..830P} Parker, E.~N.\ 1957, Physical
Review, 107, 830



\bibitem[Panaitescu et al.(1999)]{1999ApJ...522L.105P} Panaitescu, A.,
Spada, M., \& M{\'e}sz{\'a}ros, P.\ 1999, \apjl, 522, L105

\bibitem[Pe'er
\& Waxman(2005)]{2005ApJ...628..857P} Pe'er, A., \& Waxman, E.\ 2005, \apj, 628, 857


\bibitem[Perna et al.(2006)]{2006ApJ...636L..29P} Perna, R., Armitage,
P.~J., \& Zhang, B.\ 2006, \apjl, 636, L29


\bibitem[Proga
\& Zhang(2006)]{2006MNRAS.370L..61P} Proga, D., \& Zhang, B.\ 2006, \mnras, 370, L61


\bibitem[Ramirez-Ruiz
\& Fenimore(2000)]{2000ApJ...539..712R} Ramirez-Ruiz, E., \& Fenimore, E.~E.\ 2000, \apj, 539, 712


\bibitem[Romano et
al.(2006)]{2006A&A...450...59R} Romano, P., Moretti, A., Banat, P.~L., et al.\ 2006, \aap, 450, 59


\bibitem[Shibata
\& Magara(2011)]{2011LRSP....8....6S} Shibata, K., \& Magara, T.\ 2011, Living Reviews in Solar Physics, 8, 6



\bibitem[Sweet(1958)]{1958IAUS....6..123S} Sweet, P.~A.\ 1958,
Electromagnetic Phenomena in Cosmical Physics, 6, 123



\bibitem[Swenson
\& Roming(2014)]{2014ApJ...788...30S} Swenson, C.~A., \& Roming, P.~W.~A.\ 2014, \apj, 788, 30


\bibitem[Uhm
\& Zhang(2015)]{2015arXiv150903296U} Uhm, Z.~L., \& Zhang, B.\ 2015, arXiv:1509.03296

\bibitem[Waxman (1997)]{1997PhRvL..78.2292W} Waxman, E., 1997, ApJ, 491, L19


\bibitem[Wang
\& Dai(2013)]{2013NatPh...9..465W} Wang, F.~Y., \& Dai, Z.~G.\ 2013, Nature Physics, 9, 465


\bibitem[Wang et al.(2015)]{2015ApJS..216....8W} Wang, F.~Y., Dai, Z.~G.,
Yi, S.~X., \& Xi, S.~Q.\ 2015, \apjs, 216, 8


\bibitem[Wheatland et al.(1998)]{1998ApJ...509..448W} Wheatland, M.~S.,
Sturrock, P.~A., \& McTiernan, J.~M.\ 1998, \apj, 509, 448

\bibitem[Wu et al.(2013)]{2013ApJ...767L..36W} Wu, X.-F., Hou, S.-J.,
\& Lei, W.-H.\ 2013, \apjl, 767, L36


\bibitem[Yi et al.(2013)]{2013ApJ...776..120Y} Yi, S.-X., Wu, X.-F.,
\& Dai, Z.-G.\ 2013, \apj, 776, 120


\bibitem[Yi et al.(2015)]{2015ApJ...807...92Y} Yi, S.-X., Wu, X.-F., Wang,
F.-Y., \& Dai, Z.-G.\ 2015a, \apj, 807, 92

\bibitem[Yi et al.(2015)] {2015} Yi, S. X., Lei, W. H., Zhang, B., Dai, Z. G., Wu, X. F., \& Liang, E. W. 2015b, submitted


\bibitem[Zhang et al.(2006)]{2006ApJ...642..354Z} Zhang, B., Fan, Y.~Z.,
Dyks, J., et al.\ 2006, \apj, 642, 354


\bibitem[Zhang
\& Yan(2011)]{2011ApJ...726...90Z} Zhang, B., \& Yan, H.\ 2011, \apj, 726, 90



\bibitem[Zhang(2007)]{2007HiA....14...41Z} Zhang, S.~N.\ 2007, Highlights
of Astronomy, 14, 41


\end{thebibliography}
\end{document}